\documentclass[twocolumn,           
               showpacs,            
               preprintnumbers,     
               aps,                 
               prd,          	    
               letterpaper,             
               groupaddress,      
               nofootinbib,         
               tightenlines,        
               floats,floatfix      
               ]{revtex4-1}

\usepackage{graphicx}
\usepackage{dcolumn}
\usepackage{bm}
\usepackage{amsmath}
\usepackage{epstopdf}

\begin{document}

\title{Unified description of the dynamics of quintessential scalar
  fields}

\author{L. Arturo Ure\~na-L\'opez}%
 \email{lurena@fisica.ugto.mx}
\affiliation{%
Departamento de F\'isica, DCI, Campus Le\'on, Universidad de
Guanajuato, 37150, Le\'on, Guanajuato, M\'exico.}

\date{\today}

\begin{abstract}
Using the dynamical system approach, we describe the general dynamics
of cosmological scalar fields in terms of critical points and
heteroclinic lines. It is found that critical points describe the
initial and final states of the scalar field dynamics, but that
heteroclinic lines give a more complete description of the
evolution in between the critical points. In particular, the
heteroclinic line that departs from the (saddle) critical point of
perfect fluid-domination is the representative path in phase space of
quintessence fields that may be viable dark energy candidates. We also
discuss the attractor properties of the heteroclinic lines, and their
importance for the description of thawing and freezing fields.
\end{abstract}

\maketitle

\section{\label{sec:level1}Introduction}
The dynamics of scalar fields has been under a careful scrutiny since
their involvement as strong candidates in dark energy
models; those models useful for dark energy are generically dubbed as
quintessence fields, in order to distinguish them from purely
inflationary scalar
fields\cite{Ratra:1987rm,Zlatev:1998tr,Li:2011sd,*Copeland:2006wr,*Tsujikawa:2010sc}. There
seems to be a consensus in that there are two types of extreme
behaviors for quintessence fields: thawing and
freezing\cite{Caldwell:2005tm}. Thawing fields have at the beginning a
behavior quite similar to that of a cosmological constant, .i.e., its
energy density is almost constant and its equation of state (EOS) is
close to $w \simeq -1$. On the other side, freezing fields have an EOS
far away from the cosmological constant behavior, but that evolves
towards $w \simeq -1$ at late times.

More recently, there have been serious attempts to show that the
dynamics of cosmological scalar fields should proceed under the action
of some simple guidelines determined both by the scalar field
equations of motion and the presence of other background fields needed
for a complete description of the evolution of the
Universe\cite{delaMacorra:1999ff,Linder:2006sv,Scherrer:2007pu,Cahn:2008gk,Dutta:2008qn,Dutta:2011ik}. In
a first approximation, we may think that the quintessence field is in
a slow-roll regime similar to that of inflationary models, as in the
two cases the main purpose is to provide of an accelerating expansion
of the Universe. However, contrary to the inflationary case, the
quintessence field, if really existent, has not yet dominated the
expansion of the Universe and must have co-existed with other dominant
components in the past. This co-existence should have made its imprint
on the early evolution of the quintessence field and determined the
present conditions for an accelerating Universe. These simple
arguments suffice to show that the quintessence dynamics cannot be
simply described by the slow-roll formalism of
inflation\cite{Cahn:2008gk}. In fact, it has been shown that thawing
and freezing fields have quite distinct behaviors at early times.

In this paper, we shall show that general guidelines exist for the
evolution of cosmological scalar fields, which are given in terms of
the phase space structure of appropriately chosen scalar field
variables. These scalar field variables are simply the kinetic and
potential contributions of the scalar field to the total energy
density of the Universe which, to our knowledge, were first defined in
the seminal paper of Copeland, Liddle and Wands in
Ref.\cite{Copeland:1997et} for the simple model of a quintessence
field endowed with an exponential potential.

Our claim is that dynamical \emph{attractor trajectories} exists in
the form of the \emph{heteroclinic trajectories} of the phase space,
and that these special trajectories give an unified description and
guidance for the dynamics of general scalar field models
(see\cite{Reyes:2010zzb,UrenaLopez:2007vz,*Kiselev:2008zm,*Kiselev:2009xm}
for a similar result for inflationary fields). We are then taking a
step further from the standard approach in cosmological dynamical
systems: it is not enough to search for the critical (fixed) points of
the system, but also important is to determine the special
trajectories that connect physically relevant critical points. In
passing by, we shall argue that the phase space of the kinetic and
potential variables is the appropriate arena for the description of
the dynamics of general scalar field models, irrespective of the
number of dimensions the true phase space needs to fully describe the
evolution of the quintessence field.

A brief description of the paper is as follows. In
Sec.~\ref{sec:dynamical-system} we present the general equations of
motion and review the properties of the dynamical system corresponding
to the case of an exponential scalar field potential; we will
emphasize the role played by critical points and heteroclinic lines in
the determination of the scalar field dynamics and the structure of
the phase space. In Sec.~\ref{sec:dynam-gener-scal} we extend the
description of the exponential potential to more general cases, again
in terms of critical points and heteroclinic lines. In particular, we
give a separate description for the dynamics of thawing and freezing
fields, and provide of simple explanations for the expected
  behavior of quintessence fields. We present numerical results in
Sec.~\ref{sec:numerical-examples} to support the qualitative
description given in previous sections. Finally,
Sec.~\ref{sec:conclusions} is devoted to conclusions.

\section{\label{sec:dynamical-system}Dynamical system}
Let us consider a flat Friedmann-Robertson-Walker (FRW) Universe that
contains a perfect fluid with a barotropic equation of state (EOS)
$p_\gamma = (\gamma -1) \rho_\gamma$, where $p_\gamma$ and
$\rho_\gamma$ are its pressure and its energy density,
respectively. The known values of the EOS are $\gamma=1$ for
pressureless matter (dust), and $\gamma =4/3$ for a relativistic fluid
(radiation), but in general $0 < \gamma < 2$. There is also a scalar
field $\phi$ which is endowed with a scalar potential $V(\phi)$. The
calculations that follow will not depend upon the particular form
$V(\phi)$, but we will focus our attention on potentials that can show
a useful quintessence behavior.

The equations of motion, which together form the well known
Einstein-Klein-Gordon (EKG) system of equations, read
\begin{subequations}
  \label{eq:motion}
  \begin{eqnarray}
    \dot{H} &=& \frac{\kappa^2}{2} (\gamma \rho_\gamma + \dot{\phi}^2
    ) \, , \label{eq:motiona} \\
    \dot{\rho}_\gamma &=& -3H \gamma \rho_\gamma \,
    , \label{eq:motionb} \\
    \ddot{\phi} &=& - 3 H \dot{\phi} - \partial_\phi V \,
    , \label{eq:motionc}
  \end{eqnarray} 
\end{subequations}
Where $H = \dot{a}/a$ is the Hubble parameter and a dot means
derivative with respect to cosmic time. In addition, one has the
Friedmann constraint
\begin{equation}
  \label{eq:friedmann}
  H^2 = \frac{\kappa^2}{3} \left( \rho_\gamma + \frac{1}{2}
    \dot{\phi}^2 + V(\phi) \right) \, ,
\end{equation}
with $\kappa^2 = 8\pi G$, and $G$ is  Newton's constant. The energy
density of a homogeneous scalar field is $\rho_\phi = (1/2)
\dot{\phi}^2 + V(\phi)$, whereas its pressure is $p_\phi = (1/2)
\dot{\phi}^2 - V(\phi)$.

We follow here the dynamical system approach of Copeland, Liddle and
Wands in Ref.\cite{Copeland:1997et}. If we define the new variables
\begin{equation}
  \label{eq:dynxy}
  x \equiv \frac{\kappa \dot{\phi}}{\sqrt{6}H} \, , \quad y \equiv
  \frac{\kappa \sqrt{V}}{\sqrt{3}H} \, ,
\end{equation}
then the Klein-Gordon (KG) equation of motion of the scalar
field~(\ref{eq:motionc}) can be written as a dynamical system of the
form
\begin{subequations}
  \label{eq:dynamical}
  \begin{eqnarray}
    x^\prime &=& - 3 x + \lambda \sqrt{\frac{3}{2}} y^2 + \frac{3}{2}
    x \left[ 2x^2 + \gamma (1 -x^2 - y^2 ) \right] \,
    , \label{eq:dynamicala} \\
    y^\prime &=& -\lambda \sqrt{\frac{3}{2}} x y + \frac{3}{2} y
    \left[ 2x^2 + \gamma (1 -x^2 - y^2 ) \right] \,
    , \label{eq:dynamicalb}
  \end{eqnarray}
\end{subequations}
where we have defined the roll parameter $\lambda = - \partial_\phi
V/(\kappa V)$, and a prime denotes a derivative with respect to the
$e$-folding number $N \equiv \ln(a)$. The Friedmann
constraint~(\ref{eq:friedmann}) now reads
\begin{equation}
  \label{eq:friedmann2}
  \frac{\kappa^2 \rho_\gamma}{3 H^2} + x^2 + y^2 = 1 \, ,
\end{equation}
and finally Eq.~(\ref{eq:motiona}) reads
\begin{equation}
  \label{eq:2ndfriedmann}
  \frac{\dot{H}}{H^2} = - \frac{3}{2} \left[ 2x^2 + \gamma (1 -x^2 -
    y^2 ) \right] \, .
\end{equation}
Notice that it is not necessary to consider the equation of motion of
the perfect fluid~(\ref{eq:motionb}), as we can use the Friedmann
constraint~(\ref{eq:friedmann2}) to determine $\rho_\gamma$ once we
have found a solution for $x$ and $y$ from
Eqs.~(\ref{eq:dynamical}). We can also write the (non-constant)
barotropic EOS $\gamma_\phi$, and the density parameter $\Omega_\phi$
of the scalar field, as
\begin{equation}
  \label{eq:sfvariables}
  \gamma_\phi = \frac{p_\phi + \rho_\phi}{\rho_\phi} = \frac{2 x^2}{x^2
  + y^2} \, , \quad \Omega_\phi = \frac{\kappa^2 \rho_\phi}{3 H^2} =
x^2 + y^2 \, .
\end{equation}

It is well known that the Eqs.~(\ref{eq:dynamical}) form an autonomous
system if $\lambda = \mathrm{const.}$, which is the case of an
exponential scalar potential of the form $V(\phi) = V_0 e^{-\lambda
  \kappa \phi}$. Exponential potentials are well understood dynamical
systems, as infered from the many papers that have studied its
dynamical properties in different cosmological
settings\cite{Copeland:1997et,Barreiro:1999zs,*vandenHoogen:1999qq,*Heard:2002dr,*Guo:2003eu,*Neupane:2003cs,*Collinucci:2004iw,*UrenaLopez:2005zd,*EscamillaRivera:2010zz,*EscamillaRivera:2010py,*Obregon:2010nt,Hartong:2006rt},
see also \cite{Li:2011sd,*Copeland:2006wr,*Tsujikawa:2010sc} for a
modern and general review on scalar field models.

For a general scalar potential, it is necessary to
consider an equation of motion for $\lambda$ itself, which in general
terms is given
by\cite{Fang:2008fw,Scherrer:2007pu,Chongchitnan:2007eb}
\begin{equation}
  \label{eq:lambda}
  \lambda^\prime = - \sqrt{6} x \lambda^2 (\Gamma -1) \, ,
\end{equation}
where $\Gamma = V \partial_{\phi \phi} V/ (\partial_\phi V)^2$ is the
so-called tracker parameter\cite{Zlatev:1998tr}. For those cases in which
Eq.~(\ref{eq:lambda}) can be written in terms only of variables
$x,y,\lambda$, then Eqs.~(\ref{eq:dynamical}) and~(\ref{eq:lambda})
together form a three dimensional autonomous system. This was indeed
the case thoroughly studied in Ref.\cite{Fang:2008fw}.

It is not our intention to do the same here, but we shall rather take
a more general approach that includes even those cases in which
Eq.~(\ref{eq:lambda}) cannot be written in a closed form, and the
equations of motion of higher derivatives of the potential would have
to be taken into account. 

\subsection{\label{sec:crit-points-stab}Critical points and stability}
\begin{table*}[htp]
\caption{\label{tab:critical} Critical points $(x_0,y_0)$ of the
  dynamical system~(\ref{eq:dynamical}) in the case of an exponential
  potential of the form $V(\phi) = V_0 e^{-\lambda \kappa \phi}$,
  taken from Ref.\cite{Copeland:1997et}. See text below for more
  details.}
\begin{ruledtabular}
\begin{tabular}{c c c c c c c}
Label & $x_0$  & $y_0$ & Existence  & Stability & $\Omega_\phi$ &
$\gamma_\phi$ \\ 
\hline
A     & $0$  & $0$ & $\forall \lambda$ and $\gamma$ & Saddle point for
$0 < \gamma < 2 $ & 0 & Undefined \\ 
B     & $1$   & $0$ & $\forall \lambda$ and $\gamma$ & Unstable node
for $\lambda < \sqrt{6}$ & 1 & 2 \\ 
& & & & Saddle point for $\lambda > \sqrt{6}$ & & \\
C     & $-1$   & $0$ & $\forall \lambda$ and $\gamma$ & Unstable node
for $\lambda > - \sqrt{6}$ & 1 & 2 \\ 
& & & & Saddle point for $\lambda < -\sqrt{6}$ & & \\
D     & $\lambda/\sqrt{6}$  & $\sqrt{1-\lambda^2/6}$  & $\lambda^2 <
6$   & Stable node for $\lambda^2 < 3\gamma$ & 1 & $\lambda^2/3$ \\ 
 & & & & Saddle point for $3\gamma < \lambda^2 < 6$ & & \\
E     & $\sqrt{3/2} \gamma /\lambda$ & $\sqrt{3(2-\gamma) \gamma /
  2\lambda^2}$ & $\lambda^2 > 3\gamma$ & Stable node for $3\gamma <
\lambda^2 < 24 \gamma^2/(9\gamma - 2)$ & $3\gamma/\lambda^2$ &
$\gamma$ \\
   &       &         &         & Stable spiral for $\lambda^2 > 24
   \gamma^2 /(9\gamma - 2)$ & &
\end{tabular}
\end{ruledtabular}
\end{table*}

We start by recalling the properties of the \emph{critical points}, see
also\cite{wiggins2010}, $(x_0,y_0)$, those for which
$x^\prime(x_0,y_0) = 0 = y^\prime (x_0,y_0)$, of the 2-dim
system~(\ref{eq:dynamical}) in the case of an exponential potential,
$\lambda = \mathrm{const.}$, which are shown in
Table~\ref{tab:critical}. Critical points are also called \emph{fixed
  points}, because they represent in themselves a solution of the
dynamical system that appears on the phase space $(x,y)$ as the single
point $(x_0,y_0)$. There are in general four types of critical points,
whose existence and stability conditions depend upon the values of the
perfect fluid EOS $\gamma$ and the roll parameter $\lambda$.

To study the stability properties of the critical points, one
considers small perturbations of the form $x = x_0 + u$ and $y = y_0 +
v$, where $(u,v)$ are the perturbation variables. By taking a first
order expansion of the full equations of motion~(\ref{eq:dynamical})
around a given critical point $(x_0,y_0)$, the linear equations of
motion can be written as
\begin{equation}
  \label{eq:linear}
  \left(
    \begin{array}{c}
      u^\prime \\
      v^\prime
    \end{array}
  \right) = \mathcal{M}   \left(
    \begin{array}{c}
      u \\
      v
    \end{array}
  \right) \, , \quad \mathcal{M} =   \left(
    \begin{array}{cc}
      \partial_x (x^\prime) & \partial_y (x^\prime) \\
      \partial_x (y^\prime) & \partial_y (y^\prime)
    \end{array}
  \right)_{x_0,y_0} \, .
\end{equation}
The solution of the perturbation variables is formally given by
$\mathbf{u} = e^{\mathcal{M}N} \mathbf{u}_i$, where the exponential
term is itself a matrix that can be written in terms of the
eigenvalues $m$ and eigenvectors $\bar{\mu}$ of the stability
matrix $\mathcal{M}$. Actually, the general solution for a given
critical point is of the form
\begin{equation}
  \label{eq:perturbation}
  \left(
    \begin{array}{c}
      u \\
      v
    \end{array}
  \right) = C_1 \bar{\mu}_1 e^{m_1 N} +  C_2 \bar{\mu}_2 e^{m_2 N}
\end{equation}
where $C_1,C_2$ are arbitrary constants related to the initial
conditions.

If all eigenvalues have negative (positive) real part, $\mathrm{Re}(m)
<0 (>0)$, then the corresponding critical point is called stable
(unstable); any imaginary part in the eigenvalues will add an
oscillatory feature to the solutions and they are further label as
spirals. In the case there is one eigenvalue with a positive real part
and another one with a negative real part, then the critical point is
called a saddle.

For each one of the eigenvalues there is one eigenvector, and hence
there are two eigenvectors for each critical point. Each eigenvector
$\bar{\mu}$ determines a principal direction around a critical point
along which the solution is purely given by the corresponding
eigenfunction $e^{m N}$.

\subsection{\label{sec:phase-space-struct}Phase space structure}
Before we proceed further, we analyze the structure of the phase space
in the case of an exponential potential. It is usually assumed that
one should only take care of the upper part of phase space $y > 0$ for
expanding Universes. However, the fact is that the upper and lower
parts of the phase space are separated regions, each one enclosed by
boder lines called as separatrix. In more strict terms, a separatrix is
not but the \emph{heteroclinic line} that connnects two different
critical points.

Heteroclinic lines depart from an unstable or saddle point along its
unstable principal direction, and arrives at another stable or saddle
point along its stable principal direction. In the case of
Eqs.~(\ref{eq:dynamical}), we can identify four or five heteroclinic
lines, depending on the values of $\lambda$, as the latter determines
the number and nature of the critical points, see
Table~\ref{tab:critical}.

\begin{figure}[!htbp]
\includegraphics[width=0.49\textwidth]{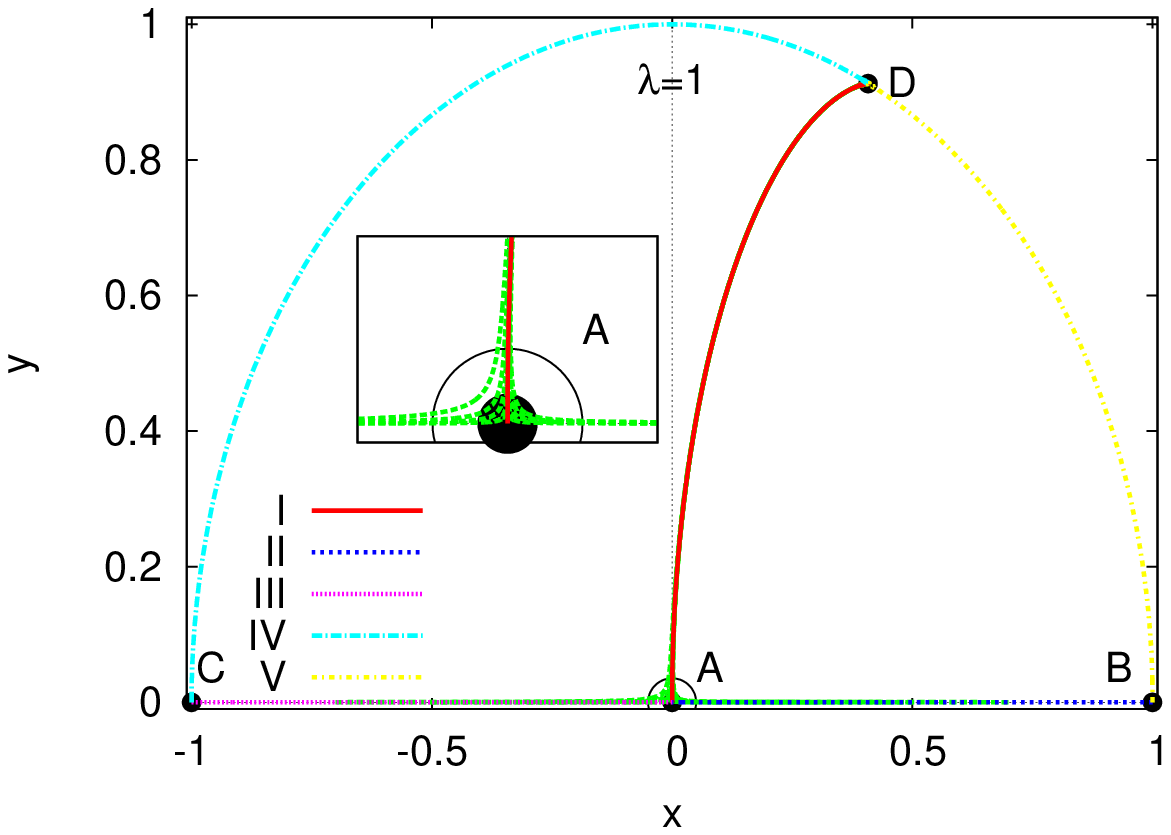}
\includegraphics[width=0.49\textwidth]{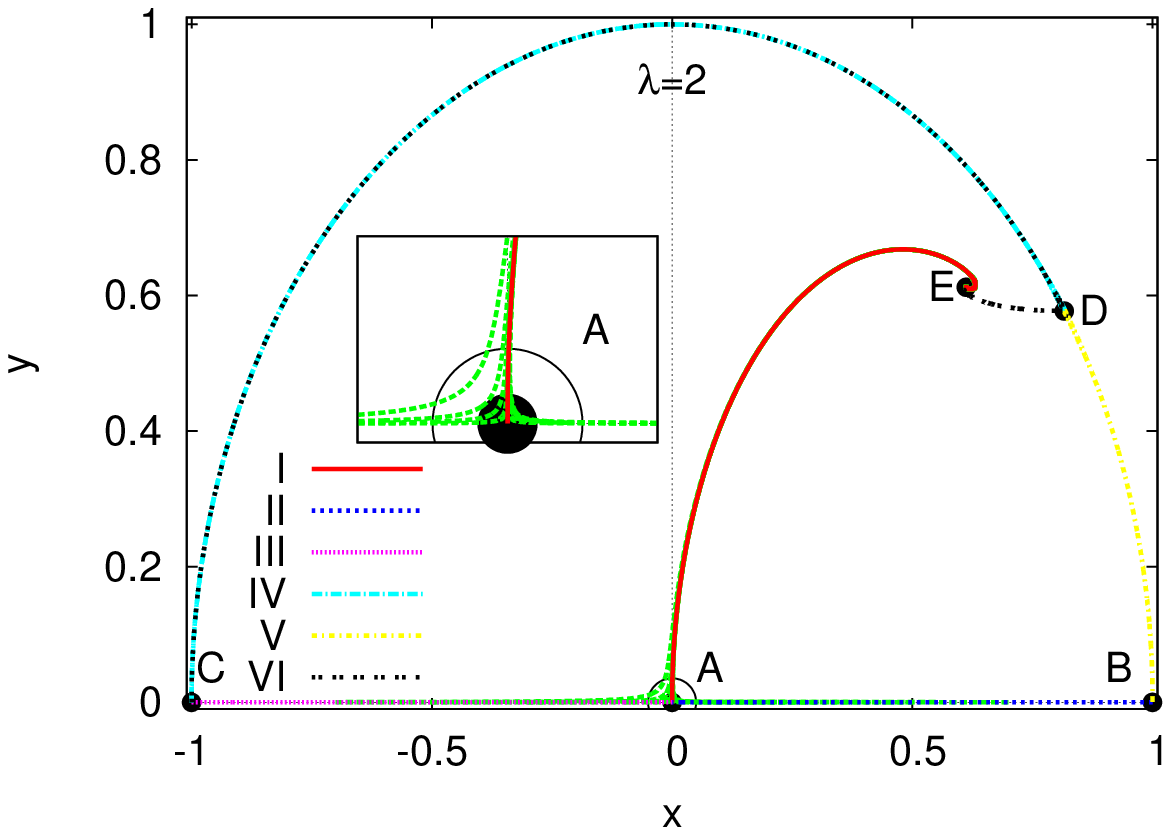}
\includegraphics[width=0.49\textwidth]{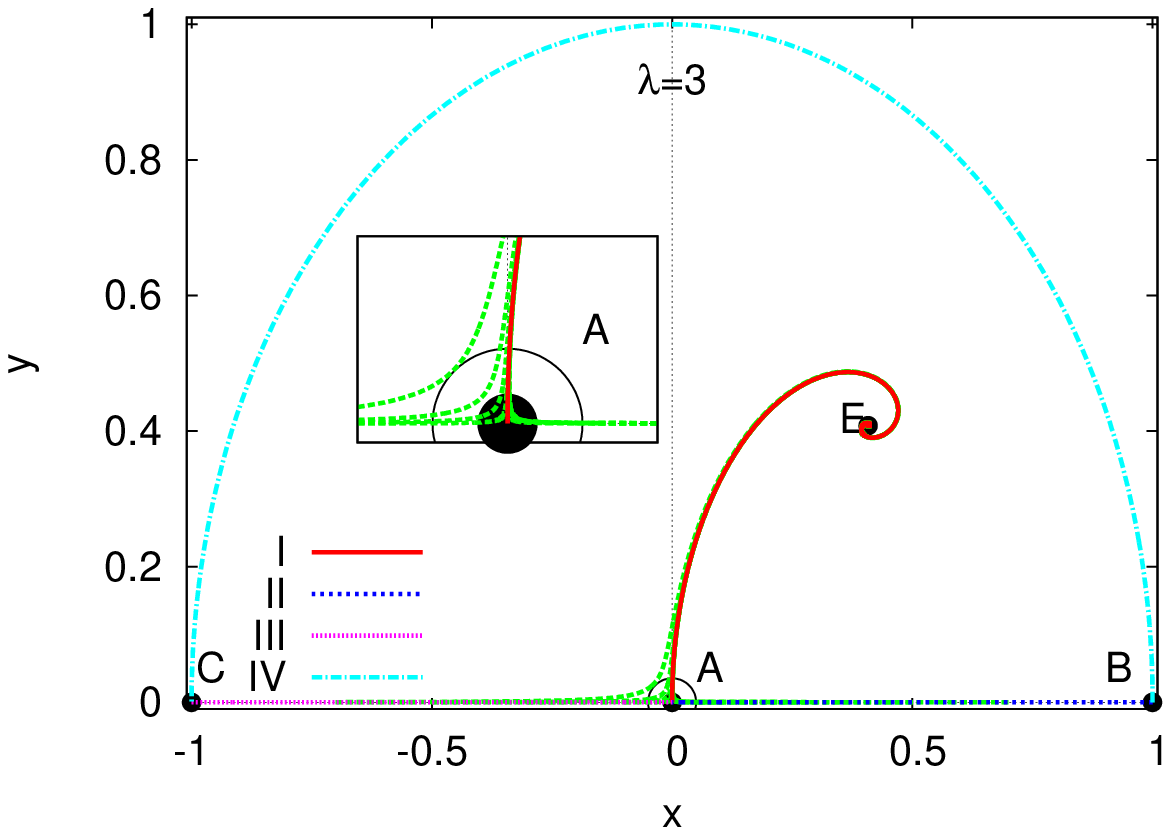}
\caption{\label{fig:structure}Structure of the phase space according to the
dynamical system~(\ref{eq:dynamical}) in the case of $\lambda =
\mathrm{const.}$ We show the critical points of
Table~\ref{tab:critical} and the heteroclinic trajectories described
in Sec.~\ref{sec:dynamical-system} that join them in each case, see
text for details. Notice that the phase space is split in different
disjoint regions for which the heteroclinic lines act as
separatrices. Also shown are different trajectories in phase space
  obtained as numerical solutions of the dynamical system
  ~(\ref{eq:dynamical}), that depart from different points with $y_i =
  10^{-4}$. It can be seen that all trajectories are \emph{focused}
  around the heteroclinic line I that connects the saddle point A (see
  the insets) with the corresponding stable point D or E in each
  case. The semicircle around the origin of coordinates corresponds to
  $\Omega_\phi = x^2 + y^2 = (0.05)^2$.}
\end{figure}

One can show that there are three heteroclinic trajectories that
correspond to the conditions $y=0$ and $x^2 + y^2 = 1$ (it can be
easily verified by direct substitution that these conditions are exact
solutions of the full dynamical system~(\ref{eq:dynamical})). The
heteroclinic trajectories for $(x,y=0)$ are those that depart from the
unstable kinetic-dominated points $(\pm 1,0)$ (B and C, respectively,
in Table~\ref{tab:critical}) and arrive at the saddle perfect fluid
dominated point $(0,0)$ (point A in Table~\ref{tab:critical}). These
heteroclinic lines are labeled as II and III in the plots in
Fig.~\ref{fig:structure}, and are also the separatrices of the $y > 0$
and $y<0$ regions of the 2-dim phase space. In the case $x^2 + y^2 =
1$, we can have one heteroclinic trajectory departing from the unstable
kinetic point $(-1,0)$ and arriving to the the saddle kinetic point
$(1,0)$ if $\lambda^2 > 3\gamma$ (line IV in bottom plot in
Fig.~\ref{fig:structure}); or we can have two heteroclinic lines
departing each one from any of the unstable kinetic points $(\pm 1
,0)$ and arriving to the (saddle or stable) scalar field dominated
point, corresponding to $3\gamma < \lambda^2 < 6$ or $\lambda^2 <
3\gamma$, respectively (lines IV and V in top and middle plots in
Fig.~\ref{fig:structure}). In any case, the heteroclinic lines $x^2 +
y^2 = 1$ do not allow variables $x,y$ to take values larger than
unity. This is a manifestation of spatial flatness as imposed upon our
model by the Friedmann constraint~(\ref{eq:friedmann2}). 

There is not an analytic expression for the heteroclinic line that
connects the saddle perfect fluid dominated point $(0,0)$ (A in
Table~\ref{tab:critical}) to the stable scalar field-dominated
solution (for $\lambda^2 < 3\gamma$, point D in
Table~\ref{tab:critical}) or to the scaling solution (for $\lambda^2 >
3\gamma$, point A in Table~\ref{tab:critical}), that we have labeled
as I in all cases in Fig.~\ref{fig:structure}. There is also one more
heteroclinic line that joins the saddle scalar field-dominated and the
stable scaling solutions, but it only appears for $3\gamma < \lambda^2
< 6$ (line VI in middle plot in Fig.~\ref{fig:structure}). 

As we said before, heteroclinic lines split the phase space in
separated regions that are disconnected one from each other, that is,
any trajectory must remain within the region it initially started in. There
are two separated regions for $\lambda^2 < 6 $, but only one single
bounded region if $\lambda^2 > 6$.

\subsection{\label{sec:gener-dynam-expon}General dynamics for the
  exponential case}
The general dynamics of a scalar field endowed with an exponential
potential is completely described by the properties of the critical
points and the heteroclinic lines of its phase space. Examples of
different trajectories were given in Ref.\cite{Copeland:1997et}. Here
we will focus our attention only in trajectories that are consistent
with an early perfect-fluid dominated Universe. 

In Fig.~\ref{fig:structure} we also show examples of different
trajectories departing closely from the perfect fluid domination point
A. The general behavior of these particular trajectories can be quoted
quite simply: each trajectory lies close to the heteroclinic line
that joins the saddle perfect fluid-dominated point A and the stable
critical point at hand in each case (see the heteroclinic lines I in
all plots in Fig.~\ref{fig:structure}). We can turn this quote into a
stronger statement: \emph{the general dynamics of a cosmological
  scalar field endowed with an exponential potential, is well
  described by the heteroclinic trajectory that joins the perfect
  fluid dominated point with the late time attractor present in the
  phase space}.

By studying this single trajectory, one can in principle determine the
agreement of a given model, in the exponential case, with the expected
evolution of the Universe. From the numerical point of view, the
drawing of the heteroclinic trajectory needs of very especial initial
conditions. It is not possible to find a full analytical solution for
the heteroclinic trajectory we are interested in, but at least the
behavior of the trajectory close to the origin of coordinates can be
easily determined as follows.

We first write the linear equations of motion that are used to study
the stability of critical points in autonomous systems. For the case
of the perfect fluid-dominated point A, the linearization of the
system~(\ref{eq:dynamical}) has the explicit solution
  \begin{equation}
    \label{eq:perfectfluid}
      \left(
    \begin{array}{c}
      x \\
      y
    \end{array}
  \right) = x_i \left(
    \begin{array}{c}
      1 \\
      0
    \end{array}
  \right) e^{-3(2-\gamma) N/2} + y_i \left( 
    \begin{array}{c}
      0 \\
      1
    \end{array}
  \right) e^{3\gamma N/2} \, ,
  \end{equation}
where we used the original phase space variables $(x,y)$ for the
perturbations $(u,v)$, which is very convenient as the perturbation
procedure is made around the origin of coordinates. Here, $x_i$ and
$y_i$ refer to the initial values of the phase space variables.

For both cases of a relativistic and a pressureless fluid, the first
eigenvalue is negative definite, $m_1 = -3(2-\gamma)/2 < 0$, and then
the kinetic energy of the scalar field must \emph{decrease} for
trajectories nearby the critical point A. On the other hand, the
second eigenvalue is positive definite, $m_2 = 3\gamma/2 > 0$, and
then the potential energy must \emph{increase}. In terms of the
dynamical system jargon, one says that the critical point A is a
saddle point and its normal directions, represented by the
eigenvectors of the stability matrix $\mathcal{M}$, see
Eq.~(\ref{eq:linear}), coincide with those of the phase space axes,
see Eq.~(\ref{eq:perfectfluid}).

We have learned then that the kinetic energy of the scalar field
decreases if the latter is subdominant with respect to the perfect
fluid, which is the expected situation in the early Universe. This
also means that the kinetic variable $x$ must be smaller than the
potential one $y$. If we now make a second order expansion of
Eq.~(\ref{eq:dynamicala}), but still a first order expansion of
Eq.~(\ref{eq:dynamicalb}), around the same critical point A, we find
\begin{subequations}
  \label{eq:secondorder}
  \begin{eqnarray}
    x^\prime &=& -\frac{3(2-\gamma)}{2} x + \lambda \sqrt{\frac{3}{2}}
    \, y^2  \, , \\
    y^\prime &=& \frac{3\gamma}{2} y \, ,
  \end{eqnarray}
\end{subequations}
whose general solutions, under the assumption that $\lambda$ is
constant, are
\begin{subequations}
  \label{eq:secondsolution}
  \begin{eqnarray}
    x(N) &=&  x_i e^{-3(2-\gamma) N/2} + \sqrt{\frac{2}{3}}
    \frac{\lambda}{(2+\gamma)} y^2 \, , \label{eq:secondsolutiona} \\
    y(N) &=& y_i e^{3\gamma N/2} \, . \label{eq:secondsolutionb}
  \end{eqnarray}
\end{subequations}
It can be seen that, if we neglect the decaying solution, the kinetic
energy evolves proportionally to the square of the potential
energy, namely
\begin{equation}
  \label{eq:thawingconstraint}
  x(y) = \frac{2}{(2+\gamma)} \frac{\lambda}{\sqrt{6}} y^2 \, .
\end{equation}
This is in good agreement with our assumption that the kinetic
perturbation $x$ is one order of magnitude smaller than potential
perturbation $y$. For future reference, we shall call
Eq.~(\ref{eq:thawingconstraint}) the \emph{thawing constraint}. Notice
that Eq.~(\ref{eq:thawingconstraint}) has the same form as the famous
slow-roll condition
\begin{equation}
  \label{eq:slow-roll}
  3 H \dot{\phi} \simeq - \partial_\phi V \quad \Rightarrow \quad x(y)
  = \frac{\lambda}{\sqrt{6}} y^2 \, ,
\end{equation}
except for the correction induced by the presence of the perfect fluid
through its EOS $\gamma$.

We have found the early behavior of the heteroclinic line that departs
from the perfect fluid-dominated point, in the phase space the
trajectory corresponds to the loci of a rotated parabola. Such a
behavior can be easily seen in all lines I in Fig.~\ref{fig:structure};
actually, the heteroclinic trajectories in  Fig.~\ref{fig:structure}
were found from the numerical solution of Eqs.~(\ref{eq:dynamical})
with initial conditions subjected to Eq.~(\ref{eq:thawingconstraint}).

Eq.~(\ref{eq:thawingconstraint}) can be written in the more
recognizable terms of the flow parameter $F \equiv \gamma_\phi
/(\Omega_\phi \lambda^2)$, defined by Cahn, de Putter, and Linder in
Ref.\cite{Cahn:2008gk}. If we take the thawing
constraint~(\ref{eq:thawingconstraint}) and the definitions in
Eq.~(\ref{eq:sfvariables}), we find that
\begin{equation}
  \label{eq:flow00}
  F \simeq \frac{4}{3(2+\gamma)^2} \, ,
\end{equation}
and then we recover the known value $F = 4/27$ if the dominant perfect
fluid is pressureless matter $(\gamma =1)$, whereas $F=3/25$ if it is
a relativistic fluid $(\gamma =4/3)$. In passing by, we note that the
value of the flow parameter under the terms of the slow-roll
condition~(\ref{eq:slow-roll}) is $F=1/3$. The asymptotic values of
some other quantities around the point $(0,0)$ can also be found, as
for instance\cite{Linder:2006sv,Cahn:2008gk}:
\begin{equation}
  \frac{3H\dot{\phi}}{-\partial_\phi V} \simeq
  \frac{2}{2+\gamma} \, , \quad \frac{\ddot{\phi}}{\partial_\phi V}
  \simeq - \frac{\gamma}{2+\gamma} \, .
\end{equation}

As we have said before, critical points in the exponential case are
also fixed points, and then they seem to have more relevance that the
trajectories themselves on phase space. This is true in the case of
early and late time attractors, but particular trajectories are
important if in addition one is interested in solutions which are in
agreement with the stringent constraints arising from the radiation
and matter dominated eras of the early Universe.

In Fig.~\ref{fig:structure} we show that trajectories with initial
values close to the perfect fluid-dominated point A follow a quite
similar path to that of the heteroclinic trajectory that connects
point A with any of the stable fixed points at hand. The seemingly
attractor character, or focusing property, of this heteroclinic line
comes from a combination of the saddle properties of point A and the
attractor character of the ending stable point. In the following
sections, we will argue that this nice property is preserved in the
case of general scalar field potentials.

\section{\label{sec:dynam-gener-scal}Dynamics of general scalar field
  models}
In the case of an exponential potential, critical points are truly
equilibrium fixed points. For the case $\lambda$ is also a variable,
those points are not fixed, but they still represent, for a given
value of $\lambda$, the points at which the 2-dim phase space velocity
$(x^\prime,y^\prime)$ vanishes. In other words, the velocity field on
the 2-dim phase space $(x,y)$ has the same structure than that in the
exponential case, but now the location of the critical points can
change as $\lambda$ evolves.

If $\lambda$ changes slowly, the critical points can be treated as if
they were approximately fixed points; this feature is particularly
useful in the case of quintessence fields, as one expects slow
variations in the roll parameter $\lambda$ to have a solution with an
accelerating expansion of the Universe at late times when the scalar
field is the dominant component. However, the same can happen at early
times, when the scalar field is subdominant with respect to the
radiation and matter components. Hence, we expect the early and late
time evolution of general scalar field models to proceed as in the
case of the exponential potential, even though the intermediate-time
dynamics of the scalar field may be more involved due to the evolution
of the roll parameter $\lambda$. 

\subsection{\label{sec:early-dynam-thaw}Early dynamics of thawing and
  freezing scalar field solutions}
From the dynamical system perspective, $\lambda$ plays an important
role in determining the existence and stability properties of two
important critical points: the scalar field dominated solution D and
the scaling solution E, see Table~\ref{tab:critical}. As we shall
show, some aspects of the early and late time evolution of scalar
fields can be understood if those two critical points are taken into
account.

For simplicity, let us start with the assumption that $\lambda \ll
3\gamma$, and that $\lambda$ is a monotonic growing function of $N$
(this condition is not crucial for the argument, but it helps to
visualize the description below). Thus, the scaling solution does not
exist, and the only critical solution on the 2-dim phase space is that
of scalar field domination, point D. If $\lambda$ changes slowly and
keeps a value below $3\gamma$, the 2-dim velocity flow will focus the
trajectories to that point, and then we shall recover a scalar
field-dominated and accelerating Universe as long as $\lambda <
\sqrt{2}$ at the present time.

Notice that we have just described the evolution expected for
thawing solutions, in which the EOS starts very close to the
cosmological constant value, $\gamma_\phi \simeq 0$, the flow
parameter is $F = 4/27$, and the field gains kinetic energy as its
evolution proceeds in the form suggested by
Eqs.~(\ref{eq:thawingconstraint}). As the value of $\lambda$ grows,
the scalar field-dominated point D \emph{moves clockwise} and so does
the phase space trajectory (see, for instance, Fig.~\ref{fig:powerlaw}
below).

Another extreme case is that in which $\lambda^2 \gg 3 \gamma$, and
$\lambda$ is a monotonic decreasing function of $N$ (as before, this
condition is not crucial for the argument, but it also helps to
visualize the description below). For this case, the scalar
field-dominated solution D does not exist, but the scaling solution E
comes into play. For large values of $\lambda$, this latter critical
point is located very close to the perfect fluid domination one, point
A, see Table~\ref{tab:critical}. Again, if the variation of $\lambda$
is slow enough, the field follows the heteroclinic trajectory that
departs from the perfect fluid-dominated solution, hooks up with the
scaling solution and follows it as the value of $\lambda$ changes. It
should be stressed out here that we are referring to the same
heteroclinic line described above for thawing fields and which is
partially described by Eq.~(\ref{eq:thawingconstraint}).

The scaling point E \emph{moves outwards} from the origin in phase space
if $\lambda$ decreases, and then $\gamma_\phi \simeq \gamma$ throughout
this period; if at some point $\lambda^2 < 3\gamma$, then the scaling
solution disappears and the scalar field then looks for the scalar
field-dominated solution D. Contrary to the previous case, the
trajectory of the scalar field \emph{moves counterclockwise} in the
phase space (see, for instance, Fig.~\ref{fig:inverse} below). While
the scalar field sits down on the scaling solution, we easily infer
that the flow parameter is $F = 1/3$, and then we have just described
the evolution of freezing tracker fields.

We have seen that for a tracker solution we need the value of
$\lambda$ to allow the appearance of the scaling solution close to the
perfect fluid dominating solution, which is assured by asking
$\lambda^2 \gg 3 \gamma$. For intermediate values $\lambda \sim
3\gamma$, the scaling solution may not exist or, if it exists, it is
located far away from the origin of the phase space. Thus, the
behavior in the proximity of the origin must be that of the thawing
solution. However, notice that a purely thawing potential, for which
$\lambda$ grows, will never be able to accelerate the expansion of the
Universe at late times if initially $\lambda \sim 3\gamma$. 

If the same is assumed for a purely freezing potential, for which
$\lambda$ decreases, then the scalar field will anyway show a thawing
behavior, because the scaling solution is never at hand for it to
catch it up. The freezing potential has the chance to accelerate the
expansion of the Universe, if $\lambda$ decreases fast enough as to
fulfill the condition $\lambda < \sqrt{2}$ before the present time.

\subsection{\label{sec:geom-interpr-flow}Geometric interpretation of
  the Flow parameter}
There is a simple geometrical interpretation of the flow parameter $F$
in Eq.~(\ref{eq:flow00}) in terms of trajectories in phase space. As
we have said before, what we have called the thawing
solution~(\ref{eq:thawingconstraint}) indeed corresponds to the
initial part of the heteroclinic trajectory that joins the (saddle)
perfect fluid-dominated solution A to any of the stable solutions at
hand. Most trajectories seem then to be \emph{focused} towards the
heteroclinic trajectory, and then the adjective \emph{flow} is well
deserved for the parameter $F$.

In the case of freezing behavior, we have shown that trajectories
first engage with the thawing solution and then hook up with the
scaling solution; in consequence, the flow parameter takes on the
freezing value $F=1/3$. The flow now refers to the fact that solutions
are following the scaling solution as it evolves outwards from the
phase space origin. The focusing of the trajectories is achieved via
the stability properties of the scaling solution.

\subsection{\label{sec:tracker-theorem}The tracker theorem}
The nice attractor features of freezing fields at early times come
basically from its tracker nature, a term first applied to scalar
fields in Ref.\cite{Zlatev:1998tr}. In this paper, the authors
established a theorem for the necessary conditions for a scalar field
to show tracking behavior; in simple terms, one needs at early times
the tracker parameter to be nearly constant and to satisfy the
condition $(\Gamma - 1) > 0$. One can though understand the tracker
conditions in more simple terms by using the dynamical system
equations~(\ref{eq:dynamical}).

We have seen that the attractor property of freezing fields arises
because one sets up large initial values of the roll parameter
$\lambda$, so that the scaling solution E is a critical point of the
2-dim phase space velocity $(x^\prime,y^\prime)$. The larger the value of
$\lambda$, the easier is for the scalar field to catch up the scaling
solution. From this point of view, the tracker condition just refers
to large initial values of the roll parameter that allow the
appearance of the scaling solution E. (As a curiosity, one can see
that the attractor character of the tracker solution discussed in
Ref.\cite{Zlatev:1998tr} actually coincides with the analysis made in
Ref.\cite{Copeland:1997et} for the scaling solution.)

However, one can go back to the usual tracker conditions if we take a
look at the evolution equation of the roll parameter,
Eq.~(\ref{eq:lambda}). It can be seen that large initial values of the
roll parameter are only allowed for scalar field models for which
$\lambda$ is a growing function if we go backwards in time; that is,
we need $(\Gamma -1) > 0$. What we do not need is the slow variation
of the tracker parameter, because, as we shall show in the numerical
experiments of Sec.~\ref{sec:numerical-examples}, the very condition
$\lambda \gg 1$ leads to $x \to 0$. Hence, it is not the tracker
parameter but the vanishing of the kinetic energy the responsible for
the slow-variation of the roll parameter at early times, see
Eq.~(\ref{eq:lambda}). This enhances the possibilities of the scalar
field to catch up the scaling solution, as the corresponding critical
point on the 2-dim phase space $(x,y)$ moves adiabatically.

Therefore, the tracker theorem can be restated in simpler terms as
follows. \emph{The tracker property appears for any scalar field model
  in which the roll parameter $\lambda$ is capable of taking on large
  initial values in the early Universe}. It is easier to look at the
expected evolution of $\lambda$, than to explore the more involved
tracker parameter $\Gamma$, to determine the tracker properties of a
given scalar field model.

\subsection{\label{sec:late-time-dynamics}Late time dynamics}
The late time dynamics of the quintessence fields will depend upon its
thawing or freezing behavior. In the case of thawing fields, we expect
that they slowly deviate from the cosmological constant values, and
that $\lambda$ moves from small to large values. If a thawing field is
currently accelerating the expansion of the Universe, then $\lambda <
\sqrt{2}$ today. It may be the case that $\lambda$ takes even larger
values in the future, so that the quintessence field is not able to
accelerate the cosmic expansion and can even provoke the appearance
again of the scaling solution E on the phase space if at some point
$\lambda^2 > 3 \gamma$.

As for freezing fields, they start with large values of $\lambda$, but
the latter has to migrate to smaller values $\lambda < \sqrt{2}$ so
that the inflationary scalar field-dominated solution D appears on
time. If this were also the trend for the future, we can only expect
that a freezing field will only come closer to the cosmological
constant behavior, i.e. $x \to 0$  and $y \to 1$, as $\lambda \to 0$.

Whether the field is thawing or freezing, its capability for providing
an accelerating solution nowadays can be characterized in terms of
$\lambda$, as it is only necessary that the (inflationary) scalar field
dominated solution D becomes alive. Hence, nowadays $\lambda$ should
be of the order of $\lambda^2 \simeq 3(1+w_{\phi,0})$.

Even if the early dynamics of a scalar field resembles very much that
of an exponential potential, the late-time dynamics can be more
involved and strongly depends upon the explicit form of the scalar
field potential\cite{delaMacorra:1999ff,Fang:2008fw}. Nonetheless, an
easy description can be given for those cases in which
Eqs.~(\ref{eq:dynamical}) and~(\ref{eq:lambda}) together form a closed
and autonomous system of equations\cite{Fang:2008fw}. For these cases,
the phase space is truly 3-dimensional, $(x,y,\lambda)$, and we will
refer to it hereafter as the augmented phase space. Notice that the
heteroclinic lines I, II, and III described in
Sec.~\ref{sec:phase-space-struct} retain its character for any value
of $\lambda$; in geometric terms, we can see that the augmented phase
space occupies, in general, the interior region of a unitary cylinder
whose axis lies along the $\lambda$-direction.

If there is a value $\lambda_\ast$ such that $[\Gamma (\lambda_\ast)
-1] =0$, then there exist generalized scalar field-dominated and
scaling solutions in the augmented phase space $(x,y,\lambda)$. The
properties of these critical fixed points are quite similar to their
exponential counterparts in Table~\ref{tab:critical} in that one only
needs to make the replacement $\lambda \to \lambda_\ast$. Moreover,
there also exists the generalized perfect fluid-dominated solution,
which exists for any value $\lambda$ and is always a saddle point.

Whenever the generalized scalar field dominated D and scaling E
solutions exist, there can be a \emph{unified description} of the
dynamics in terms again of the heteroclinic line that joins the
perfect fluid dominated point A to the stable point available in the
augmented phase space. As in the case of the exponential potential,
the general solution for a given model is then characterized by such
heteroclinic line. Even more, in our numerical experiments with
different potentials in Sec.~\ref{sec:numerical-examples}, we will
find that the early description of the heteroclinic points is also
given by the thawing behavior in
Eq.~(\ref{eq:thawingconstraint}). This was to be expected, as the
evolution of $\lambda$ depends upon the second order perturbations of
$x$, see Eq.~(\ref{eq:lambda}), and then most of the motion around the
perfect fluid-dominated solution must proceed at $\lambda \simeq
\mathrm{const.}$

There is though a non-trivial difference with respect to the
exponential case: the de Sitter point $(0,1)$ exists as a late time
attractor if $\lambda \to 0$ and $[\Gamma(0)-1] > 0$. That is, for any
scalar field model in which the roll parameter vanishes at late times
the final fate is the cosmological constant
case\cite{delaMacorra:1999ff,Fang:2008fw}.

\section{\label{sec:numerical-examples}Numerical examples}
We now present some numerical solutions of the EKG equations in cases
in which Eqs.~(\ref{eq:dynamical}) and~(\ref{eq:lambda}) form a closed
and autonomous system of equations. This is done for numerical
purposes, as in other more general cases we may need to write a full
hierarchy of equations for higher derivatives of the scalar field
potential. As we have discussed before, it will be only necessary to
look at the properties of the 2-dim phase space $(x,y)$ to have an
ample picture about the dynamics of scalar fields.

\subsection{\label{sec:monot-freez-case}Monotonic freezing case}
As a typical example for a freezing model we take the tracker models
$V(\phi)= M^{4+n} \phi^{-n}$, with $n >0$, for which the roll
parameter and the tracker parameter are, respectively, $\lambda =
n/(\kappa \phi)$ and $\Gamma - 1 = 1/n > 0$; notice that this is a
\emph{monotonic freezing} case, as the latter condition on the tracker
parameter means that the roll parameter is an always \emph{decreasing}
function. In fact, the equation of motion of the roll parameter is
\begin{equation}
  \label{eq:lambdafreezing}
    \lambda^\prime = - \frac{\sqrt{6}}{n} x \lambda^2 \, ,
\end{equation}
and then it is one of those fortunate cases in which the equations of
motion~(\ref{eq:dynamical}) and~(\ref{eq:lambdafreezing}) together
form an autonomous system of equations, which is able to represent the
full dynamics of the original EKG system.

According to the phase space analysis in previous sections, this type
of models have the following \emph{fixed} points in the augmented
phase space on the plane $\lambda = 0$: perfect fluid-dominated point
A, kinetic dominated points B and C, and the de Sitter point
$(0,1)$. In addition, points D and E are also critical, but not fixed,
points at different slices with $\lambda = \mathrm{const.}$ Even
though we cannot apply here the stability analysis of
Ref.\cite{Fang:2008fw}, we expect the de Sitter point to be a stable
point: once the field is rolling down its potential we have $x>0$, and
then $\lambda$ should be an ever decreasing function according to
Eq.~(\ref{eq:lambdafreezing}). In this form, the system must approach
the de Sitter point asymptotically.

\begin{figure}[t]
\includegraphics[width=0.49\textwidth]{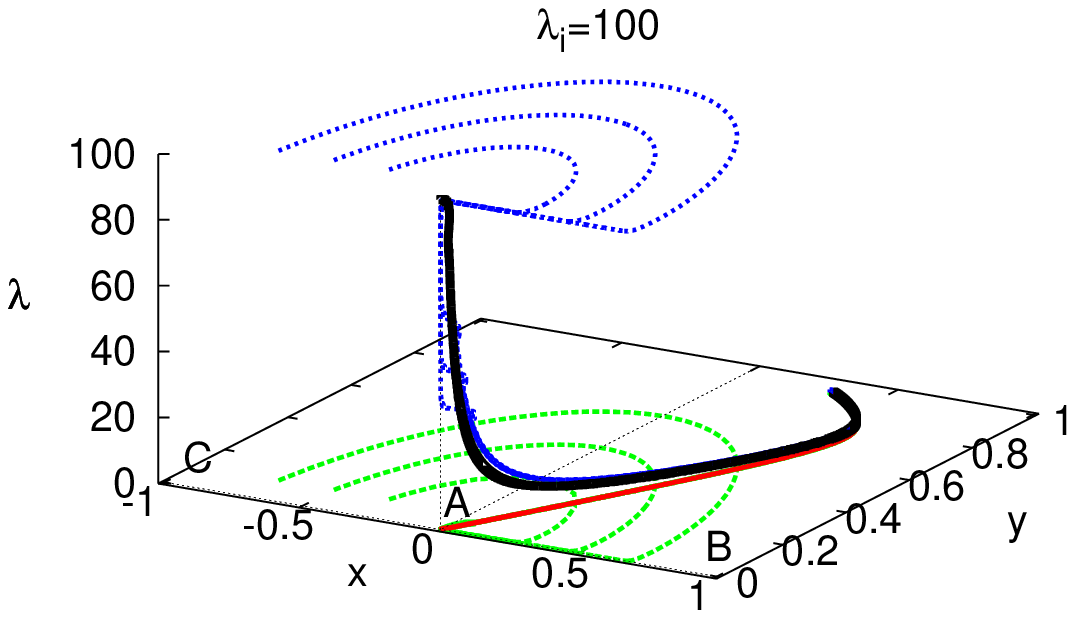}
\caption{\label{fig:3Dinverse}Numerical solutions of
  Eqs.~(\ref{eq:dynamical}) and~(\ref{eq:lambdafreezing}), for
  numerical convenience we chose $n=50$. Shown are the trajectories
  (black-solid and blue-dotted lines) in the 3D phase space
  $(x,y,\lambda)$ for different initial conditions of the
  system~(\ref{eq:dynamical}) and~(\ref{eq:lambdafreezing}). The
  corresponding projections onto the $(x,y)$ plane are also shown
  (red-solid and green-dashed lines); see the text and
  Fig.~\ref{fig:inverse} for a comprehensive description of the
  dynamics on the $xy$-plane.}
\end{figure}

\begin{figure}[!htbp]
\includegraphics[width=0.49\textwidth]{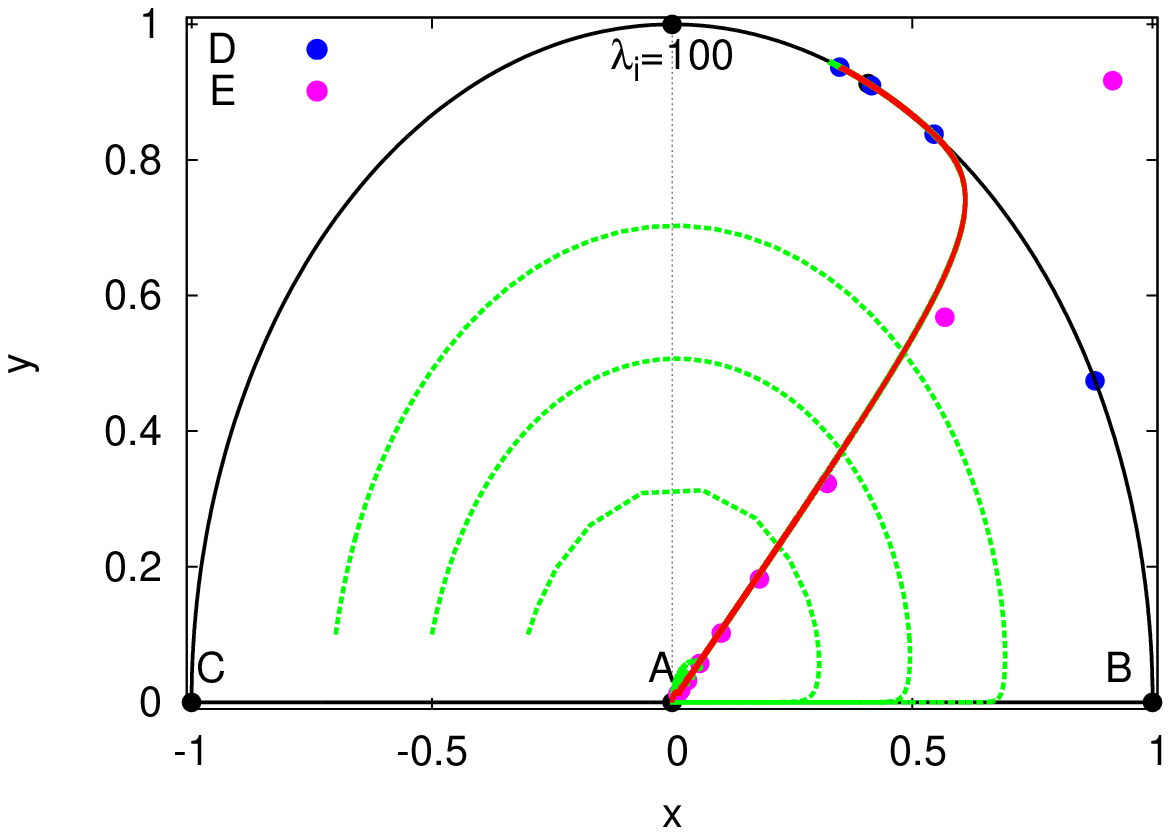}
\includegraphics[width=0.49\textwidth]{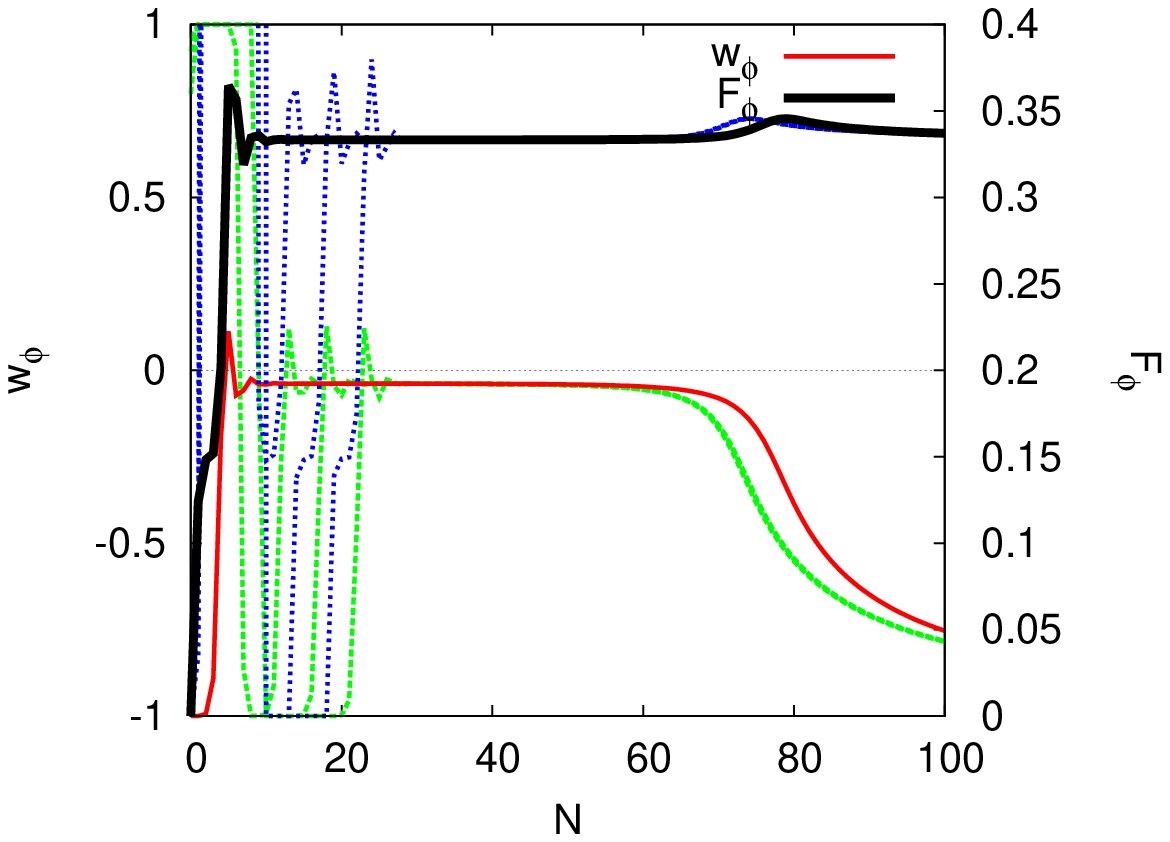}
\caption{\label{fig:inverse}Numerical solutions of
  Eqs.~(\ref{eq:dynamical}) and~(\ref{eq:lambdafreezing}), which
  together form an autonomous dynamical system; for numerical
  convenience we chose $n=50$. (Top) Trajectories of the numerical
  solutions on the 2-dim $(x,y)$ plane. The dots denote the positions
  of the critical points D and E, see Table~\ref{tab:critical}, as
  $\lambda$ evolves. All the solutions initially have
  $\lambda_i = 100$. In particular, the (red solid) line is the
  heteroclinic line that joins the perfect fluid-dominated point A
  with the de Sitter point $(0,1)$. (Bottom) The evolution of the
  scalar field EOS $w_\phi$ (red-solid and green-dashed lines) and the
  flow parameter $F$ (black-solid and blue-dotted lines) for the same
  numerical solutions as in the top figure. Both parameters show a
  tracker intermediate behavior for which $w_\phi \simeq
  \mathrm{const.}$ and $F = 1/3$. At late time, the EOS takes on
  negative values as the solution evolves towards the de Sitter point
  and enters a pure slow-roll regime with $F \to 1/3$ again, see
  Eq.~(\ref{eq:slow-roll}).}
\end{figure}

Numerical solutions of the set of Eqs.~(\ref{eq:dynamical})
and~(\ref{eq:lambdafreezing}) are given in Figs.~\ref{fig:3Dinverse}
for the 3-dim phase space $(x,y,\lambda)$, and in
Fig.~\ref{fig:inverse} for the 2-dim phase space $(x,y)$, for
$n=50$. A detailed description of the behavior of the different curves
is as follows. The attractor trajectory is the heteroclinic line that
joins the saddle point A $(0,0)$ with the stable de Sitter point
$(0,1)$\footnote{We refer here to a 2-dim heteroclinic line
  that connects the critical points $(0,0)$ and $(0,1)$, even though
  the true heteroclinic line exists in the augmented 3-dim space
  $(x,y,\lambda)$, and connects the points $(0,0,0)$ and
  $(0,1,0)$\cite{Fang:2008fw}, see also Fig.~\ref{fig:3Dinverse}. We
  will use the 2-dim nomenclature as it fits well what we want to
  describe on the 2-dim phase space.}; the behavior of this
heteroclinic line in the neighborhood of point A is given by the
thawing solution~(\ref{eq:thawingconstraint}). Because of the large
initial value of the roll parameter, $\lambda_i =100$, the scaling
solution, point E, is present. The heteroclinic trajectory hooks up
with the scaling point, and the latter moves outwards as the value of
$\lambda$ decreases. Once $\lambda < \sqrt{6}$, there also appears the
scalar field dominated solution, point D, which starts to move
counterclockwise on the unitary circumference as $\lambda$ continues
decreasing. Both critical points merge once $\lambda = 3\gamma$ and
the scaling point formally disappears, and the motion of the
heteroclinic line is now guided by the still moving point D until they
reach together the de Sitter point.

The description we have given for the heteroclinic line suffices to
understand the general behavior of inverse power-law potentials. To
show the attractor feature of the heteroclinic line, we have also
solved the equations of motion for diverse initial conditions
$(x_i,y_i)$ but keeping the same $\lambda_i=100$. It is clear that the
curves follow the heteroclinic line at late times, as expected. But
for this type of potentials we notice a peculiar early behavior
denoted by the clockwise semi-circumferences. This behavior is easily
explained if we write the equations of motion~(\ref{eq:dynamical}) in
the limit $\lambda \gg 1$, and then
\begin{equation}
  \label{eq:largelambda}
  x^\prime \simeq  \lambda \sqrt{\frac{3}{2}} y^2 \, , \quad y^\prime
  \simeq  - \lambda \sqrt{\frac{3}{2}} xy \, .
\end{equation}
It can be shown that, under these equations, the scalar field density
parameter~(\ref{eq:sfvariables}) is a conserved quantity,
i.e. $\Omega_\phi = x^2 + y^2 = \mathrm{const.}$, and then the phase
space variables should move clockwise on circumferences until they
lose enough potential energy ($y \to 0$) and the scalar field energy
is dominated by its kinetic component (there is no other option,
anyway, as the variable $y$ cannot become negative because of the
heteroclinic separatrix $(x,y=0)$). The clockwise motion is explained
by the fact that the field prefers to move downhill its potential, $x
> 0$, rather than otherwise. We finally note that the circular motion
occurs quite rapidly, an estimation can be made from
Eqs.~(\ref{eq:largelambda}), and then the $e$-fold interval for the
circular motion is $\Delta N \sim 1/\lambda$.

In Fig.~\ref{fig:inverse} we also make a translation of the scalar field
dynamics in terms of the scalar field EOS $w_\phi = \gamma_\phi -1$
(see Eq.~(\ref{eq:sfvariables})) and the flow parameter $F$ (see
Eq.~(\ref{eq:flow00})). The scalar field EOS reaches a tracker regime
that almost perfectly traces the background EOS, which is $\gamma=1$
for the numerical examples; this is actually the tracker behavior
described in detail in Ref.\cite{Zlatev:1998tr}. The small difference
between the two EOS means that the scalar field redshifts a bit slower
than the background field. This difference, even if small, is
important for the later overtaking of the background dynamics by the
scalar field. The value of $w_\phi$ asymptotically approaches the de
Sitter value, $w_\phi \to -1$.

As for the flow parameter, its value in the tracker regime is the
expected one, $F = 1/3$, which is a constant value for almost the
whole evolution. There is small 'bump' at the transition point at which
the solution leaves the tracker regime, but $F_\phi \to 1/3$
asymptotically again, signaling that the scalar field is entering into
a pure slow-roll regime at late times, see Eq.~(\ref{eq:slow-roll}).

\begin{figure}[!htbp]
\includegraphics[width=0.49\textwidth]{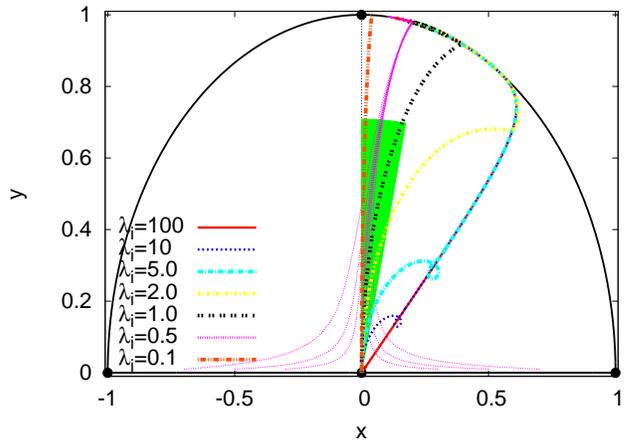}
\caption{\label{fig:inversehomoc}Heteroclinic lines departing from the
  perfect fluid-dominated point A, that were obtained from the
  numerical solutions of Eqs.~(\ref{eq:dynamical})
  and~(\ref{eq:lambdafreezing}). For each case, the initial conditions
for $x$ and $y$ were chosen according to the thawing
constraint~(\ref{eq:secondsolution}). The shaded region is a circular
arc of radius $\Omega_\phi = 0.7$ and angles corresponding to $-1 <
w_\phi < -0.94$. It can be seen that only heteroclinic lines with
$\lambda_i \leq 0.5$ may be able to provide of a physically viable
accelerating Universe.}
\end{figure}

As the heteroclinic trajectory is a strong attractor, it suffices to
scan the form of heteroclinic trajectories in phase space for different
initial values of $\lambda$ in order to find physically acceptable
accelerating solutions, as is made in Fig.~\ref{fig:inversehomoc}. We
can see that the initial value of $\lambda$ should be small enough so
that the heteroclinic line can provide of physical parameters in
agreement with cosmological observations. According to
Fig.~\ref{fig:inversehomoc}, it is necessary to have $\lambda_i < 0.5$
if we want heteroclinic trajectories to be inside the arc of radius
$\Omega_\phi = 0.7$ and angles corresponding to $-1 < w_\phi < -0.94$.

\subsection{\label{sec:monot-thaw-case}Monotonic thawing case}
To study the general behavior of thawing models, we may use the
example of a PNGB potential of the form $V(\phi) = M^4 \cos^2(\alpha
\kappa \phi)$, where $\alpha$ is a free parameter of the
model\cite{Dutta:2008qn}. The roll and tracker parameters for this
kind of potentials are
\begin{equation}
  \label{eq:thawingpars}
  \lambda = 2 \alpha \tan(\alpha \kappa \phi) \, , \quad
  \Gamma(\lambda) = - \frac{1}{2} \left( 1 +
    \frac{2\alpha^2}{\lambda^2} \right) \, .
\end{equation}
The PNGB potential is a representative case of a \emph{monotonic
thawing} case, for which the roll parameter is always an
\emph{increasing} function. However, we should remember that the PNGB
potential has a zero minimum at $\alpha \kappa \phi = \pi/2$, and it
is at this point that the roll parameter diverges, $\lambda \to
\infty$; hence, the autonomous dynamical system cannot describe the
dynamics of the scalar field model beyond this point. This is a caveat
of the dynamical system approach that appears for any model with a
zero minimum, see for instance\cite{UrenaLopez:2007vz,Matos:2009hf}. 

Some other models without a minimum in the potential can be chosen
from the examples in the specialized literature. We shall use a
generic example in which the representative features of monotonic
thawing potentials appear acutely. We write Eq.~(\ref{eq:lambda}) in
the form
\begin{equation}
  \label{eq:lambdathawing}
  \lambda^\prime = \frac{\sqrt{6}}{n} x \lambda^2 \, ,
\end{equation}
which also arises in the case of power law potentials $V \sim
\phi^n$. We have verified that other monotonic thawing potentials
display qualitatively the same generic behavior of this example
(see\cite{delaMacorra:1999ff} for some other instances).

\begin{figure}[t]
\includegraphics[width=0.49\textwidth]{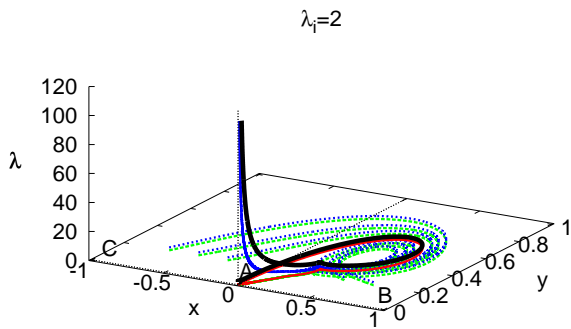}
\caption{\label{fig:3Dpowerlaw}Numerical solutions of
  Eqs.~(\ref{eq:dynamical}) and~(\ref{eq:lambdafreezing}), for
  numerical convenience we chose $n=20$. Shown are the trajectories
  (black-solid and blue-dotted lines) in the 3-dim phase space
  $(x,y,\lambda)$ for different initial conditions of the
  system~(\ref{eq:dynamical}) and~(\ref{eq:lambdathawing}). The
  corresponding projections onto the $(x,y)$ plane are also shown
  (red-solid and green-dashed lines); see the text and
  Fig.~\ref{fig:powerlaw} for a comprehensive description of the
  dynamics on the $xy$-plane.}
\end{figure}

\begin{figure}[!htbp]
\includegraphics[width=0.49\textwidth]{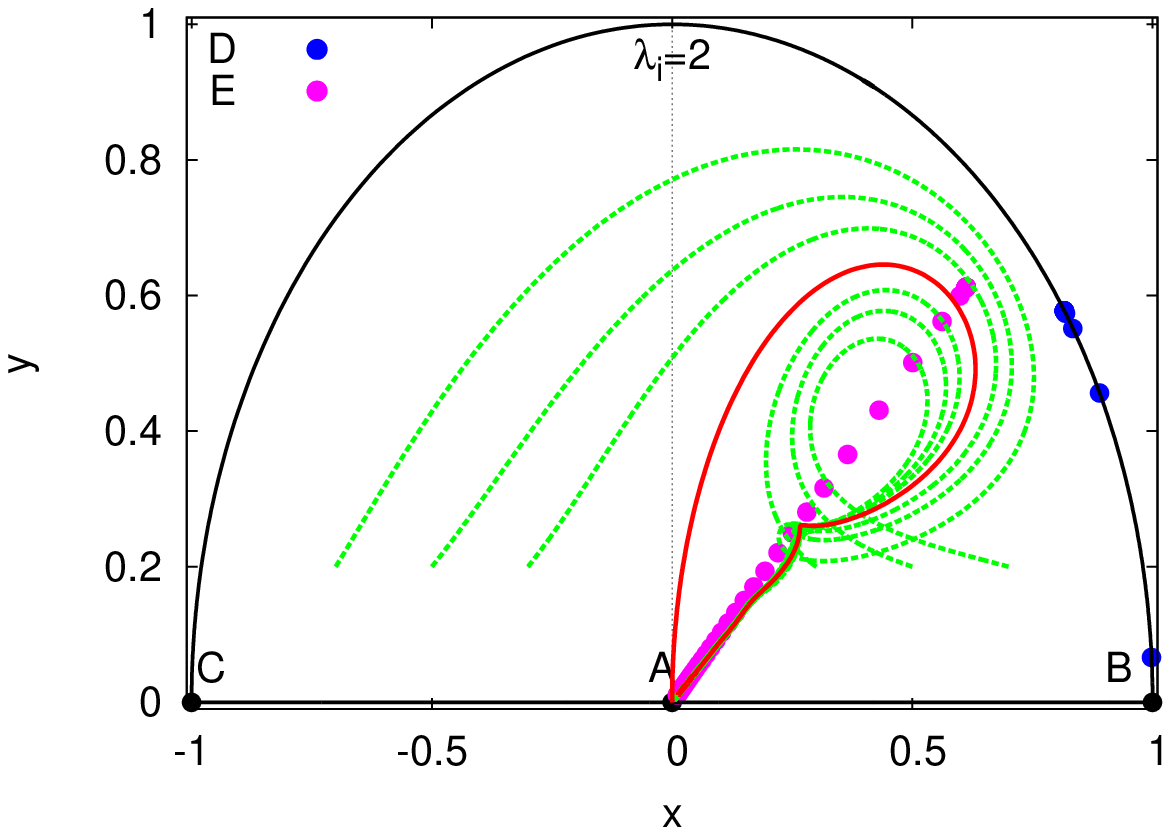}
\includegraphics[width=0.49\textwidth]{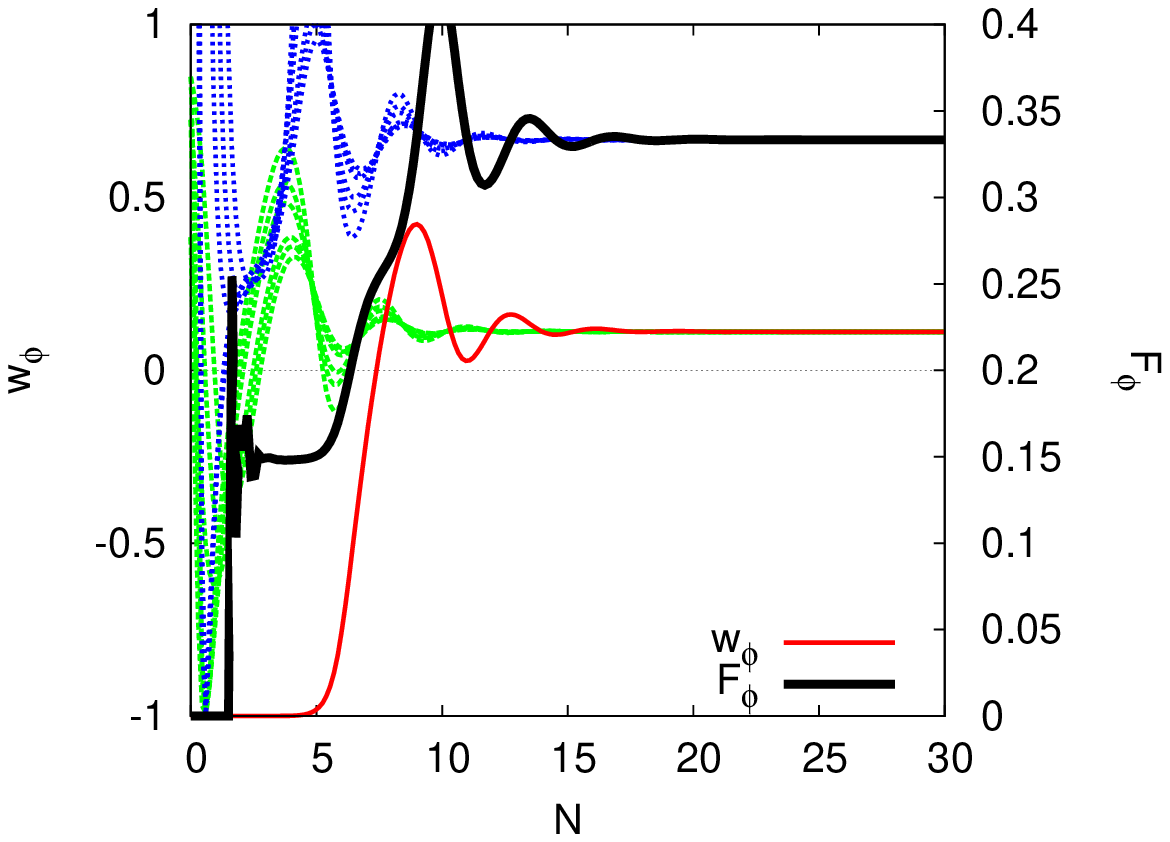}
\caption{\label{fig:powerlaw}Numerical solutions of
  Eqs.~(\ref{eq:dynamical}) and~(\ref{eq:lambdathawing}), which
  together form an autonomous dynamical system; for numerical
  convenience we chose $n=20$. (Top) Trajectories of
  the numerical solutions on the 2-dim $(x,y)$ plane. The dots
  denote the positions of the critical points D and E, see
  Table~\ref{tab:critical}, as $\lambda$ evolves. All the solutions
  initially have $\lambda_i = 2$. In particular, the (red solid) line
  is the homoclinic line that joins the perfect fluid-dominated point A
  with itself. (Bottom) The evolution of the scalar field EOS $w_\phi$
  (red-solid and green-dashed lines) and the flow parameter $F$
  (black-solid and blue-dotted lines) for the same numerical solutions
  as in the top figure. Both parameters show an early thawing behavior
  for which $w_\phi \simeq -1$ and $F = 4/27$. At late times, the EOS
  takes on positive values as the solution evolves back to the
  point $(0,0)$ and $F \to 1/3$.}
\end{figure}

As in the case of freezing models, thawing potentials has the
following \emph{fixed} points in the augmented phase space on the
plane $\lambda = 0$: perfect fluid-dominated point A, kinetic
dominated points B and C, and the de Sitter point $(0,1)$. In
addition, points D and E are also critical, but not fixed, points at
different slices with $\lambda = \mathrm{const.}$ Contrary to the
freezing case, we expect the de Sitter point to be an unstable point:
once the field is rolling down its potential we have $x>0$, and then
$\lambda$ should be an ever increasing function according to
Eq.~(\ref{eq:lambdathawing}). This latter fact has another undesirable
consequence for thawing models: there is not a fixed critical point
that can act as a late-time attractor. Thus, the thawing solution we
have found before in the neighborhood of the perfect fluid-dominated
solution, see Eq.~(\ref{eq:secondsolution}), is no longer part of a
heteroclinic line, but now we have the presence of a homoclinic line
only, that eventually connects a critical point with itself.

Numerical solutions of the set of Eqs.~(\ref{eq:dynamical})
and~(\ref{eq:lambdathawing}) are given in Fig.~\ref{fig:3Dpowerlaw}
for the 3-dim phase space $(x,y,\lambda)$, and in
Fig.~\ref{fig:powerlaw} for the 2-dim phase space $(x,y)$; in all
cases $\lambda_i =2$. The same features appear in both
Figs.~\ref{fig:inverse}, and~\ref{fig:powerlaw}, so that a quick
comparison can be made of the thawing case with the freezing case.

First of all, we cannot notice any kind of strong attractor trajectory
on the phase space at early times as it is the case in freezing
models. This was anticipated above because there is only a homoclinic
trajectory departing from and eventually arriving at the (saddle)
perfect fluid-dominated point; in practice, the homoclinic line is a
closed loop\footnote{In strict sense, what we have called a homoclinic
trajectory on the $(x,y)$ plane is not a loop, but actually an open
line in the augmented 3-dim phase space $(x,y,\lambda)$. It departs from
the point $(0,0,0)$ and arrives to the point $(0,0,\infty)$, see for
instance Fig.~\ref{fig:3Dpowerlaw}. We shall use the term homoclinic
anyway as it encloses well what we see on the $(x,y)$ plane}. The
departing behavior of the loop is given by the thawing
condition~(\ref{eq:secondsolution}), but the late time behavior is
dictated by the scaling solution E as $\lambda \to \infty$. The
homoclinic trajectory, however, do split the $(x,y)$ in two parts, as
trajectories departing at $x <0$ ($x >0$) surround the loop in the
clockwise direction by its outside (inside) part as they approach the
scaling solution E at late times.

In terms of physical variables, we notice that the scalar field EOS
for the homoclinic line departs from $w_\phi \simeq -1$, and ends up
with a constant value at late times which is a bit stiffer than the
background EOS (in this example $\gamma =1$). As for the flow
parameter, we notice the expected thawing behavior at early times, $F
= 4/27$, but also the freezing one at late times, $F=1/3$.

\begin{figure}[!htbp]
\includegraphics[width=0.49\textwidth]{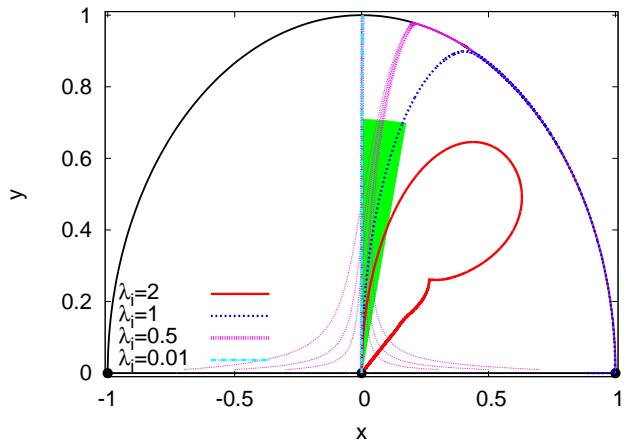}
\caption{\label{fig:powerhomoc}Homoclinic lines departing from the
  perfect fluid-dominated point A, that were obtained from the
  numerical solutions of Eqs.~(\ref{eq:dynamical})
  and~(\ref{eq:lambdathawing}). For each case, the initial conditions
  for $x$ and $y$ were chosen according to the thawing
  constraint~(\ref{eq:secondsolution}). The shaded region is a
  circular arc of radius $\Omega_\phi = 0.7$ and angles corresponding to $-1 <
w_\phi < -0.94$. It can be seen that only homoclinic lines with
$\lambda_i \leq 0.5$ may be able to provide of a physically viable
accelerating Universe.}
\end{figure}

Even though the homoclinic trajectory is not a strong attractor, it
anyway determines the way other arbitrary trajectories wander about on
the phase space. Thus, in order to search for valid accelerating
solutions, it suffices to scan the form of homoclinic trajectories in
phase space for different initial values $\lambda_i$. This is
actually shown in Fig.~\ref{fig:powerhomoc}. For small enough values
of $\lambda$, the homoclinic trajectory reaches the unitary circle
because of the presence of the scalar field-dominated point
D. Remember that this critical point exists for $\lambda^2 < 6$ and is
stable for $\lambda^2 < 2$. We also show in
Fig.~\ref{fig:powerhomoc} the homoclinic trajectories that remain
within the arc of radius $\Omega_\phi = 0.7$ and angles corresponding
to $-1 < w_\phi < 0.94$; these trajectories could provide a valid
accelerating Universe at the present time.

\section{\label{sec:conclusions}Conclusions}
We have given evidence for the existence of an unified description for
the dynamics of general scalar field models, in terms of the
heteroclinic trajectory that connects the perfect fluid-dominated
stage of the Universe with any of the late-time attractors at hand
in the phase space of the scalar field variables.

The numerical results shown in Figs.~\ref{fig:inversehomoc}
and~\ref{fig:powerhomoc} strongly suggest that, in the case of single
field models with a monotonic evolution of the roll parameter
$\lambda$, the field should start in a thawing regime far away from
the tracker one if it is to be consistent with an accelerating
Universe at late times. Actually, the tracker regime seems to spoil in
all cases the viability of quintessence models even if its early-time
attractor properties are very
appealing\cite{Zlatev:1998tr,Wang:2011bi}.

However, the early thawing behavior is a problematic one, as it
requires special initial conditions, in particular, small initial
values of the roll parameter $\lambda$. Moreover, as in the case of an
exponential potential\cite{Copeland:1997et}, the initial value of the
quintessence parameter $\Omega_\phi$ must be extremely small in order
to be consistent with cosmological constraints and to prevent
quintessence domination well before the present time. As for the late
time dynamics, the approach presented here suggests that the roll
parameter $\lambda$ is small at late times, and that its value should
be related to the present value of the dark energy EOS by $\lambda^2
\simeq 3(1+w_0)$; in other words, the derivative of the quintessence
potential must be related to the actual value of the dark energy
EOS.

It is interesting to note that the aforementioned difficulties for
thawing and freezing scalar fields are actually well known from the
case of an exponential potential\cite{Copeland:1997et}. They are
actually present in more general cases because the phase space
equations of motion preserve its form in the general case, and then
quintessence fields inherit the difficulties of the exponential case. 

There is still the possibility of having single field models with a
non-monotonic evolution of the roll parameter, in which the strong
attractor properties of freezing fields can be combined with the
appropriate late-time dynamics of thawing fields. This can in
principle be achieved with the use of very contrived potentials (see
an early example in\cite{Zlatev:1998tr}), in the case in which the
roll parameter changes stochastically, or even in the case in which
phase space dynamics needs a full hierarchy of equations of
motion\cite{Chongchitnan:2007eb}.

In any case, our general suggestion in the study of scalar field
models is still the same. One should look for the heteroclinic line
departing from the perfect fluid-domination point; this single
trajectory encloses in itself enough information to determine the
capabilities of the given quintessence field to act as a realistic
dark energy model.

\begin{acknowledgments}
I am grateful to E. Linder for enlightening discussions and useful
comments about the manuscript. I thank the Berkeley Center for
Cosmological Physics (BCCP) for its kind hospitality, and the joint
support of the Academia Mexicana de Ciencias and the United
States-Mexico Foundation for Science for a summer research stay at
BCCP. This work was partially supported by PROMEP, DAIP, and by
CONACyT M\'exico under grants 56946, 167335, and I0101/131/07 C-234/07
of the Instituto Avanzado de Cosmologia (IAC) collaboration.
\end{acknowledgments}

\bibliography{cosmosfrefs}

\begin{thebibliography}{33}%
\makeatletter
\providecommand \@ifxundefined [1]{%
 \@ifx{#1\undefined}
}%
\providecommand \@ifnum [1]{%
 \ifnum #1\expandafter \@firstoftwo
 \else \expandafter \@secondoftwo
 \fi
}%
\providecommand \@ifx [1]{%
 \ifx #1\expandafter \@firstoftwo
 \else \expandafter \@secondoftwo
 \fi
}%
\providecommand \natexlab [1]{#1}%
\providecommand \enquote  [1]{``#1''}%
\providecommand \bibnamefont  [1]{#1}%
\providecommand \bibfnamefont [1]{#1}%
\providecommand \citenamefont [1]{#1}%
\providecommand \href@noop [0]{\@secondoftwo}%
\providecommand \href [0]{\begingroup \@sanitize@url \@href}%
\providecommand \@href[1]{\@@startlink{#1}\@@href}%
\providecommand \@@href[1]{\endgroup#1\@@endlink}%
\providecommand \@sanitize@url [0]{\catcode `\\12\catcode `\$12\catcode
  `\&12\catcode `\#12\catcode `\^12\catcode `\_12\catcode `\%12\relax}%
\providecommand \@@startlink[1]{}%
\providecommand \@@endlink[0]{}%
\providecommand \url  [0]{\begingroup\@sanitize@url \@url }%
\providecommand \@url [1]{\endgroup\@href {#1}{\urlprefix }}%
\providecommand \urlprefix  [0]{URL }%
\providecommand \Eprint [0]{\href }%
\@ifxundefined \urlstyle {%
  \providecommand \doi  [0]{\begingroup \@sanitize@url \@doi}%
  \providecommand \@doi [1]{\endgroup \@@startlink {\doibase
  #1}doi:\discretionary {}{}{}#1\@@endlink }%
}{%
  \providecommand \doi  [0]{doi:\discretionary{}{}{}\begingroup
  \urlstyle{rm}\Url }%
}%
\providecommand \doibase [0]{http://dx.doi.org/}%
\providecommand \Doi [0]{\begingroup \@sanitize@url \@Doi }%
\providecommand \@Doi  [1]{\endgroup\@@startlink{\doibase#1}\@@Doi}%
\providecommand \@@Doi [1]{#1\@@endlink}%
\providecommand \selectlanguage [0]{\@gobble}%
\providecommand \bibinfo  [0]{\@secondoftwo}%
\providecommand \bibfield  [0]{\@secondoftwo}%
\providecommand \translation [1]{[#1]}%
\providecommand \BibitemOpen [0]{}%
\providecommand \bibitemStop [0]{}%
\providecommand \bibitemNoStop [0]{.\EOS\space}%
\providecommand \EOS [0]{\spacefactor3000\relax}%
\providecommand \BibitemShut  [1]{\csname bibitem#1\endcsname}%
\bibitem [{\citenamefont {Ratra}\ and\ \citenamefont
  {Peebles}(1988)}]{Ratra:1987rm}%
  \BibitemOpen
  \bibfield  {author} {\bibinfo {author} {\bibfnamefont {B.}~\bibnamefont
  {Ratra}}\ and\ \bibinfo {author} {\bibfnamefont {P.~J.~E.}\ \bibnamefont
  {Peebles}},\ }\Doi {10.1103/PhysRevD.37.3406} {\bibfield  {journal} {\bibinfo
   {journal} {Phys. Rev.},\ }\textbf {\bibinfo {volume} {D37}},\ \bibinfo
  {pages} {3406} (\bibinfo {year} {1988})}\BibitemShut {NoStop}%
\bibitem [{\citenamefont {Zlatev}\ \emph {et~al.}(1999)\citenamefont {Zlatev},
  \citenamefont {Wang},\ and\ \citenamefont {Steinhardt}}]{Zlatev:1998tr}%
  \BibitemOpen
  \bibfield  {author} {\bibinfo {author} {\bibfnamefont {I.}~\bibnamefont
  {Zlatev}}, \bibinfo {author} {\bibfnamefont {L.-M.}\ \bibnamefont {Wang}}, \
  and\ \bibinfo {author} {\bibfnamefont {P.~J.}\ \bibnamefont {Steinhardt}},\
  }\Doi {10.1103/PhysRevLett.82.896} {\bibfield  {journal} {\bibinfo  {journal}
  {Phys.Rev.Lett.},\ }\textbf {\bibinfo {volume} {82}},\ \bibinfo {pages} {896}
  (\bibinfo {year} {1999})},\ \Eprint {http://arxiv.org/abs/astro-ph/9807002}
  {arXiv:astro-ph/9807002 [astro-ph]} \BibitemShut {NoStop}%
\bibitem [{\citenamefont {Li}\ \emph {et~al.}(2011)\citenamefont {Li},
  \citenamefont {Li}, \citenamefont {Wang},\ and\ \citenamefont
  {Wang}}]{Li:2011sd}%
  \BibitemOpen
  \bibfield  {author} {\bibinfo {author} {\bibfnamefont {M.}~\bibnamefont
  {Li}}, \bibinfo {author} {\bibfnamefont {X.-D.}\ \bibnamefont {Li}}, \bibinfo
  {author} {\bibfnamefont {S.}~\bibnamefont {Wang}}, \ and\ \bibinfo {author}
  {\bibfnamefont {Y.}~\bibnamefont {Wang}},\ }\Doi {10.1088/0253-6102/56/3/24}
  {\bibfield  {journal} {\bibinfo  {journal} {Commun.Theor.Phys.},\ }\textbf
  {\bibinfo {volume} {56}},\ \bibinfo {pages} {525} (\bibinfo {year} {2011})},\
  \Eprint {http://arxiv.org/abs/1103.5870} {arXiv:1103.5870 [astro-ph.CO]}
  \BibitemShut {NoStop}%
\bibitem [{\citenamefont {Copeland}\ \emph {et~al.}(2006)\citenamefont
  {Copeland}, \citenamefont {Sami},\ and\ \citenamefont
  {Tsujikawa}}]{Copeland:2006wr}%
  \BibitemOpen
  \bibfield  {author} {\bibinfo {author} {\bibfnamefont {E.~J.}\ \bibnamefont
  {Copeland}}, \bibinfo {author} {\bibfnamefont {M.}~\bibnamefont {Sami}}, \
  and\ \bibinfo {author} {\bibfnamefont {S.}~\bibnamefont {Tsujikawa}},\ }\Doi
  {10.1142/S021827180600942X} {\bibfield  {journal} {\bibinfo  {journal} {Int.
  J. Mod. Phys.},\ }\textbf {\bibinfo {volume} {D15}},\ \bibinfo {pages} {1753}
  (\bibinfo {year} {2006})},\ \Eprint {http://arxiv.org/abs/hep-th/0603057}
  {arXiv:hep-th/0603057} \BibitemShut {NoStop}%
\bibitem [{\citenamefont {Tsujikawa}(2010)}]{Tsujikawa:2010sc}%
  \BibitemOpen
  \bibfield  {author} {\bibinfo {author} {\bibfnamefont {S.}~\bibnamefont
  {Tsujikawa}},\ }\href@noop {} { (\bibinfo {year} {2010})},\ \Eprint
  {http://arxiv.org/abs/1004.1493} {arXiv:1004.1493 [astro-ph.CO]} \BibitemShut
  {NoStop}%
\bibitem [{\citenamefont {Caldwell}\ and\ \citenamefont
  {Linder}(2005)}]{Caldwell:2005tm}%
  \BibitemOpen
  \bibfield  {author} {\bibinfo {author} {\bibfnamefont {R.}~\bibnamefont
  {Caldwell}}\ and\ \bibinfo {author} {\bibfnamefont {E.~V.}\ \bibnamefont
  {Linder}},\ }\Doi {10.1103/PhysRevLett.95.141301} {\bibfield  {journal}
  {\bibinfo  {journal} {Phys.Rev.Lett.},\ }\textbf {\bibinfo {volume} {95}},\
  \bibinfo {pages} {141301} (\bibinfo {year} {2005})},\ \Eprint
  {http://arxiv.org/abs/astro-ph/0505494} {arXiv:astro-ph/0505494 [astro-ph]}
  \BibitemShut {NoStop}%
\bibitem [{\citenamefont {de~la Macorra}\ and\ \citenamefont
  {Piccinelli}(2000)}]{delaMacorra:1999ff}%
  \BibitemOpen
  \bibfield  {author} {\bibinfo {author} {\bibfnamefont {A.}~\bibnamefont
  {de~la Macorra}}\ and\ \bibinfo {author} {\bibfnamefont {G.}~\bibnamefont
  {Piccinelli}},\ }\Doi {10.1103/PhysRevD.61.123503} {\bibfield  {journal}
  {\bibinfo  {journal} {Phys.Rev.},\ }\textbf {\bibinfo {volume} {D61}},\
  \bibinfo {pages} {123503} (\bibinfo {year} {2000})},\ \Eprint
  {http://arxiv.org/abs/hep-ph/9909459} {arXiv:hep-ph/9909459 [hep-ph]}
  \BibitemShut {NoStop}%
\bibitem [{\citenamefont {Linder}(2006)}]{Linder:2006sv}%
  \BibitemOpen
  \bibfield  {author} {\bibinfo {author} {\bibfnamefont {E.~V.}\ \bibnamefont
  {Linder}},\ }\Doi {10.1103/PhysRevD.73.063010} {\bibfield  {journal}
  {\bibinfo  {journal} {Phys.Rev.},\ }\textbf {\bibinfo {volume} {D73}},\
  \bibinfo {pages} {063010} (\bibinfo {year} {2006})},\ \Eprint
  {http://arxiv.org/abs/astro-ph/0601052} {arXiv:astro-ph/0601052 [astro-ph]}
  \BibitemShut {NoStop}%
\bibitem [{\citenamefont {Scherrer}\ and\ \citenamefont
  {Sen}(2008)}]{Scherrer:2007pu}%
  \BibitemOpen
  \bibfield  {author} {\bibinfo {author} {\bibfnamefont {R.~J.}\ \bibnamefont
  {Scherrer}}\ and\ \bibinfo {author} {\bibfnamefont {A.}~\bibnamefont {Sen}},\
  }\Doi {10.1103/PhysRevD.77.083515} {\bibfield  {journal} {\bibinfo  {journal}
  {Phys.Rev.},\ }\textbf {\bibinfo {volume} {D77}},\ \bibinfo {pages} {083515}
  (\bibinfo {year} {2008})},\ \Eprint {http://arxiv.org/abs/0712.3450}
  {arXiv:0712.3450 [astro-ph]} \BibitemShut {NoStop}%
\bibitem [{\citenamefont {Cahn}\ \emph {et~al.}(2008)\citenamefont {Cahn},
  \citenamefont {de~Putter},\ and\ \citenamefont {Linder}}]{Cahn:2008gk}%
  \BibitemOpen
  \bibfield  {author} {\bibinfo {author} {\bibfnamefont {R.~N.}\ \bibnamefont
  {Cahn}}, \bibinfo {author} {\bibfnamefont {R.}~\bibnamefont {de~Putter}}, \
  and\ \bibinfo {author} {\bibfnamefont {E.~V.}\ \bibnamefont {Linder}},\ }\Doi
  {10.1088/1475-7516/2008/11/015} {\bibfield  {journal} {\bibinfo  {journal}
  {JCAP},\ }\textbf {\bibinfo {volume} {0811}},\ \bibinfo {pages} {015}
  (\bibinfo {year} {2008})},\ \bibinfo {note} {* Brief entry *},\ \Eprint
  {http://arxiv.org/abs/0807.1346} {arXiv:0807.1346 [astro-ph]} \BibitemShut
  {NoStop}%
\bibitem [{\citenamefont {Dutta}\ and\ \citenamefont
  {Scherrer}(2008)}]{Dutta:2008qn}%
  \BibitemOpen
  \bibfield  {author} {\bibinfo {author} {\bibfnamefont {S.}~\bibnamefont
  {Dutta}}\ and\ \bibinfo {author} {\bibfnamefont {R.~J.}\ \bibnamefont
  {Scherrer}},\ }\Doi {10.1103/PhysRevD.78.123525} {\bibfield  {journal}
  {\bibinfo  {journal} {Phys.Rev.},\ }\textbf {\bibinfo {volume} {D78}},\
  \bibinfo {pages} {123525} (\bibinfo {year} {2008})},\ \Eprint
  {http://arxiv.org/abs/0809.4441} {arXiv:0809.4441 [astro-ph]} \BibitemShut
  {NoStop}%
\bibitem [{\citenamefont {Dutta}\ and\ \citenamefont
  {Scherrer}(2011)}]{Dutta:2011ik}%
  \BibitemOpen
  \bibfield  {author} {\bibinfo {author} {\bibfnamefont {S.}~\bibnamefont
  {Dutta}}\ and\ \bibinfo {author} {\bibfnamefont {R.~J.}\ \bibnamefont
  {Scherrer}},\ }\href@noop {} { (\bibinfo {year} {2011})},\ \Eprint
  {http://arxiv.org/abs/1106.0012} {arXiv:1106.0012 [astro-ph.CO]} \BibitemShut
  {NoStop}%
\bibitem [{\citenamefont {Copeland}\ \emph {et~al.}(1998)\citenamefont
  {Copeland}, \citenamefont {Liddle},\ and\ \citenamefont
  {Wands}}]{Copeland:1997et}%
  \BibitemOpen
  \bibfield  {author} {\bibinfo {author} {\bibfnamefont {E.~J.}\ \bibnamefont
  {Copeland}}, \bibinfo {author} {\bibfnamefont {A.~R.}\ \bibnamefont
  {Liddle}}, \ and\ \bibinfo {author} {\bibfnamefont {D.}~\bibnamefont
  {Wands}},\ }\Doi {10.1103/PhysRevD.57.4686} {\bibfield  {journal} {\bibinfo
  {journal} {Phys. Rev.},\ }\textbf {\bibinfo {volume} {D57}},\ \bibinfo
  {pages} {4686} (\bibinfo {year} {1998})},\ \Eprint
  {http://arxiv.org/abs/gr-qc/9711068} {arXiv:gr-qc/9711068} \BibitemShut
  {NoStop}%
\bibitem [{\citenamefont {Reyes-Ibarra}\ and\ \citenamefont
  {Urena-Lopez}(2010)}]{Reyes:2010zzb}%
  \BibitemOpen
  \bibfield  {author} {\bibinfo {author} {\bibfnamefont {M.~J.}\ \bibnamefont
  {Reyes-Ibarra}}\ and\ \bibinfo {author} {\bibfnamefont {L.}~\bibnamefont
  {Urena-Lopez}},\ }\Doi {10.1063/1.3473869} {\bibfield  {journal} {\bibinfo
  {journal} {AIP Conf.Proc.},\ }\textbf {\bibinfo {volume} {1256}},\ \bibinfo
  {pages} {293} (\bibinfo {year} {2010})}\BibitemShut {NoStop}%
\bibitem [{\citenamefont {Urena-Lopez}\ and\ \citenamefont
  {Reyes-Ibarra}(2009)}]{UrenaLopez:2007vz}%
  \BibitemOpen
  \bibfield  {author} {\bibinfo {author} {\bibfnamefont {L.~A.}\ \bibnamefont
  {Urena-Lopez}}\ and\ \bibinfo {author} {\bibfnamefont {M.~J.}\ \bibnamefont
  {Reyes-Ibarra}},\ }\Doi {10.1142/S0218271809014674} {\bibfield  {journal}
  {\bibinfo  {journal} {Int. J. Mod. Phys.},\ }\textbf {\bibinfo {volume}
  {D18}},\ \bibinfo {pages} {621} (\bibinfo {year} {2009})},\ \Eprint
  {http://arxiv.org/abs/0709.3996} {arXiv:0709.3996 [astro-ph]} \BibitemShut
  {NoStop}%
\bibitem [{\citenamefont {Kiselev}\ and\ \citenamefont
  {Timofeev}(2008)}]{Kiselev:2008zm}%
  \BibitemOpen
  \bibfield  {author} {\bibinfo {author} {\bibfnamefont {V.}~\bibnamefont
  {Kiselev}}\ and\ \bibinfo {author} {\bibfnamefont {S.}~\bibnamefont
  {Timofeev}},\ }\href@noop {} { (\bibinfo {year} {2008})},\ \Eprint
  {http://arxiv.org/abs/0801.2453} {arXiv:0801.2453 [gr-qc]} \BibitemShut
  {NoStop}%
\bibitem [{\citenamefont {Kiselev}\ and\ \citenamefont
  {Timofeev}(2010)}]{Kiselev:2009xm}%
  \BibitemOpen
  \bibfield  {author} {\bibinfo {author} {\bibfnamefont {V.}~\bibnamefont
  {Kiselev}}\ and\ \bibinfo {author} {\bibfnamefont {S.}~\bibnamefont
  {Timofeev}},\ }\Doi {10.1007/s10714-009-0827-5} {\bibfield  {journal}
  {\bibinfo  {journal} {Gen.Rel.Grav.},\ }\textbf {\bibinfo {volume} {42}},\
  \bibinfo {pages} {183} (\bibinfo {year} {2010})},\ \Eprint
  {http://arxiv.org/abs/0905.4353} {arXiv:0905.4353 [gr-qc]} \BibitemShut
  {NoStop}%
\bibitem [{\citenamefont {Barreiro}\ \emph {et~al.}(2000)\citenamefont
  {Barreiro}, \citenamefont {Copeland},\ and\ \citenamefont
  {Nunes}}]{Barreiro:1999zs}%
  \BibitemOpen
  \bibfield  {author} {\bibinfo {author} {\bibfnamefont {T.}~\bibnamefont
  {Barreiro}}, \bibinfo {author} {\bibfnamefont {E.~J.}\ \bibnamefont
  {Copeland}}, \ and\ \bibinfo {author} {\bibfnamefont {N.~J.}\ \bibnamefont
  {Nunes}},\ }\Doi {10.1103/PhysRevD.61.127301} {\bibfield  {journal} {\bibinfo
   {journal} {Phys. Rev.},\ }\textbf {\bibinfo {volume} {D61}},\ \bibinfo
  {pages} {127301} (\bibinfo {year} {2000})},\ \Eprint
  {http://arxiv.org/abs/astro-ph/9910214} {arXiv:astro-ph/9910214} \BibitemShut
  {NoStop}%
\bibitem [{\citenamefont {van~den Hoogen}\ \emph {et~al.}(1999)\citenamefont
  {van~den Hoogen}, \citenamefont {Coley},\ and\ \citenamefont
  {Wands}}]{vandenHoogen:1999qq}%
  \BibitemOpen
  \bibfield  {author} {\bibinfo {author} {\bibfnamefont {R.~J.}\ \bibnamefont
  {van~den Hoogen}}, \bibinfo {author} {\bibfnamefont {A.~A.}\ \bibnamefont
  {Coley}}, \ and\ \bibinfo {author} {\bibfnamefont {D.}~\bibnamefont
  {Wands}},\ }\Doi {10.1088/0264-9381/16/6/317} {\bibfield  {journal} {\bibinfo
   {journal} {Class. Quant. Grav.},\ }\textbf {\bibinfo {volume} {16}},\
  \bibinfo {pages} {1843} (\bibinfo {year} {1999})},\ \Eprint
  {http://arxiv.org/abs/gr-qc/9901014} {arXiv:gr-qc/9901014} \BibitemShut
  {NoStop}%
\bibitem [{\citenamefont {Heard}\ and\ \citenamefont
  {Wands}(2002)}]{Heard:2002dr}%
  \BibitemOpen
  \bibfield  {author} {\bibinfo {author} {\bibfnamefont {I.~P.}\ \bibnamefont
  {Heard}}\ and\ \bibinfo {author} {\bibfnamefont {D.}~\bibnamefont {Wands}},\
  }\Doi {10.1088/0264-9381/19/21/309} {\bibfield  {journal} {\bibinfo
  {journal} {Class.Quant.Grav.},\ }\textbf {\bibinfo {volume} {19}},\ \bibinfo
  {pages} {5435} (\bibinfo {year} {2002})},\ \Eprint
  {http://arxiv.org/abs/gr-qc/0206085} {arXiv:gr-qc/0206085 [gr-qc]}
  \BibitemShut {NoStop}%
\bibitem [{\citenamefont {Guo}\ \emph {et~al.}(2003)\citenamefont {Guo},
  \citenamefont {Piao},\ and\ \citenamefont {Zhang}}]{Guo:2003eu}%
  \BibitemOpen
  \bibfield  {author} {\bibinfo {author} {\bibfnamefont {Z.~K.}\ \bibnamefont
  {Guo}}, \bibinfo {author} {\bibfnamefont {Y.-S.}\ \bibnamefont {Piao}}, \
  and\ \bibinfo {author} {\bibfnamefont {Y.-Z.}\ \bibnamefont {Zhang}},\ }\Doi
  {10.1016/j.physletb.2003.06.004} {\bibfield  {journal} {\bibinfo  {journal}
  {Phys.Lett.},\ }\textbf {\bibinfo {volume} {B568}},\ \bibinfo {pages} {1}
  (\bibinfo {year} {2003})},\ \Eprint {http://arxiv.org/abs/hep-th/0304048}
  {arXiv:hep-th/0304048 [hep-th]} \BibitemShut {NoStop}%
\bibitem [{\citenamefont {Neupane}(2004)}]{Neupane:2003cs}%
  \BibitemOpen
  \bibfield  {author} {\bibinfo {author} {\bibfnamefont {I.~P.}\ \bibnamefont
  {Neupane}},\ }\Doi {10.1088/0264-9381/21/18/007} {\bibfield  {journal}
  {\bibinfo  {journal} {Class. Quant. Grav.},\ }\textbf {\bibinfo {volume}
  {21}},\ \bibinfo {pages} {4383} (\bibinfo {year} {2004})},\ \Eprint
  {http://arxiv.org/abs/hep-th/0311071} {arXiv:hep-th/0311071} \BibitemShut
  {NoStop}%
\bibitem [{\citenamefont {Collinucci}\ \emph {et~al.}(2005)\citenamefont
  {Collinucci}, \citenamefont {Nielsen},\ and\ \citenamefont
  {Van~Riet}}]{Collinucci:2004iw}%
  \BibitemOpen
  \bibfield  {author} {\bibinfo {author} {\bibfnamefont {A.}~\bibnamefont
  {Collinucci}}, \bibinfo {author} {\bibfnamefont {M.}~\bibnamefont {Nielsen}},
  \ and\ \bibinfo {author} {\bibfnamefont {T.}~\bibnamefont {Van~Riet}},\ }\Doi
  {10.1088/0264-9381/22/7/005} {\bibfield  {journal} {\bibinfo  {journal}
  {Class.Quant.Grav.},\ }\textbf {\bibinfo {volume} {22}},\ \bibinfo {pages}
  {1269} (\bibinfo {year} {2005})},\ \Eprint
  {http://arxiv.org/abs/hep-th/0407047} {arXiv:hep-th/0407047 [hep-th]}
  \BibitemShut {NoStop}%
\bibitem [{\citenamefont {Urena-Lopez}(2005)}]{UrenaLopez:2005zd}%
  \BibitemOpen
  \bibfield  {author} {\bibinfo {author} {\bibfnamefont {L.~A.}\ \bibnamefont
  {Urena-Lopez}},\ }\Doi {10.1088/1475-7516/2005/09/013} {\bibfield  {journal}
  {\bibinfo  {journal} {JCAP},\ }\textbf {\bibinfo {volume} {0509}},\ \bibinfo
  {pages} {013} (\bibinfo {year} {2005})},\ \Eprint
  {http://arxiv.org/abs/astro-ph/0507350} {arXiv:astro-ph/0507350} \BibitemShut
  {NoStop}%
\bibitem [{\citenamefont {Escamilla-Rivera}\ \emph
  {et~al.}(2010){\natexlab{a}}\citenamefont {Escamilla-Rivera}, \citenamefont
  {Obregon},\ and\ \citenamefont {Urena-Lopez}}]{EscamillaRivera:2010zz}%
  \BibitemOpen
  \bibfield  {author} {\bibinfo {author} {\bibfnamefont {C.}~\bibnamefont
  {Escamilla-Rivera}}, \bibinfo {author} {\bibfnamefont {O.}~\bibnamefont
  {Obregon}}, \ and\ \bibinfo {author} {\bibfnamefont {L.~A.}\ \bibnamefont
  {Urena-Lopez}},\ }\Doi {10.1063/1.3473863} {\bibfield  {journal} {\bibinfo
  {journal} {AIP Conf. Proc.},\ }\textbf {\bibinfo {volume} {1256}},\ \bibinfo
  {pages} {262} (\bibinfo {year} {2010}{\natexlab{a}})}\BibitemShut {NoStop}%
\bibitem [{\citenamefont {Escamilla-Rivera}\ \emph
  {et~al.}(2010){\natexlab{b}}\citenamefont {Escamilla-Rivera}, \citenamefont
  {Obregon},\ and\ \citenamefont {Urena-Lopez}}]{EscamillaRivera:2010py}%
  \BibitemOpen
  \bibfield  {author} {\bibinfo {author} {\bibfnamefont {C.}~\bibnamefont
  {Escamilla-Rivera}}, \bibinfo {author} {\bibfnamefont {O.}~\bibnamefont
  {Obregon}}, \ and\ \bibinfo {author} {\bibfnamefont {L.}~\bibnamefont
  {Urena-Lopez}},\ }\Doi {10.1088/1475-7516/2010/12/011} {\bibfield  {journal}
  {\bibinfo  {journal} {JCAP},\ }\textbf {\bibinfo {volume} {1012}},\ \bibinfo
  {pages} {011} (\bibinfo {year} {2010}{\natexlab{b}})},\ \Eprint
  {http://arxiv.org/abs/1009.4233} {arXiv:1009.4233 [gr-qc]} \BibitemShut
  {NoStop}%
\bibitem [{\citenamefont {Obregon}\ and\ \citenamefont
  {Quiros}(2011)}]{Obregon:2010nt}%
  \BibitemOpen
  \bibfield  {author} {\bibinfo {author} {\bibfnamefont {O.}~\bibnamefont
  {Obregon}}\ and\ \bibinfo {author} {\bibfnamefont {I.}~\bibnamefont
  {Quiros}},\ }\Doi {10.1103/PhysRevD.84.044005} {\bibfield  {journal}
  {\bibinfo  {journal} {Phys.Rev.},\ }\textbf {\bibinfo {volume} {D84}},\
  \bibinfo {pages} {044005} (\bibinfo {year} {2011})},\ \Eprint
  {http://arxiv.org/abs/1011.3896} {arXiv:1011.3896 [gr-qc]} \BibitemShut
  {NoStop}%
\bibitem [{\citenamefont {Hartong}\ \emph {et~al.}(2006)\citenamefont
  {Hartong}, \citenamefont {Ploegh}, \citenamefont {Van~Riet},\ and\
  \citenamefont {Westra}}]{Hartong:2006rt}%
  \BibitemOpen
  \bibfield  {author} {\bibinfo {author} {\bibfnamefont {J.}~\bibnamefont
  {Hartong}}, \bibinfo {author} {\bibfnamefont {A.}~\bibnamefont {Ploegh}},
  \bibinfo {author} {\bibfnamefont {T.}~\bibnamefont {Van~Riet}}, \ and\
  \bibinfo {author} {\bibfnamefont {D.~B.}\ \bibnamefont {Westra}},\ }\Doi
  {10.1088/0264-9381/23/14/003} {\bibfield  {journal} {\bibinfo  {journal}
  {Class. Quant. Grav.},\ }\textbf {\bibinfo {volume} {23}},\ \bibinfo {pages}
  {4593} (\bibinfo {year} {2006})},\ \Eprint
  {http://arxiv.org/abs/gr-qc/0602077} {arXiv:gr-qc/0602077} \BibitemShut
  {NoStop}%
\bibitem [{\citenamefont {Fang}\ \emph {et~al.}(2009)\citenamefont {Fang},
  \citenamefont {Li}, \citenamefont {Zhang},\ and\ \citenamefont
  {Lu}}]{Fang:2008fw}%
  \BibitemOpen
  \bibfield  {author} {\bibinfo {author} {\bibfnamefont {W.}~\bibnamefont
  {Fang}}, \bibinfo {author} {\bibfnamefont {Y.}~\bibnamefont {Li}}, \bibinfo
  {author} {\bibfnamefont {K.}~\bibnamefont {Zhang}}, \ and\ \bibinfo {author}
  {\bibfnamefont {H.-Q.}\ \bibnamefont {Lu}},\ }\Doi
  {10.1088/0264-9381/26/15/155005} {\bibfield  {journal} {\bibinfo  {journal}
  {Class.Quant.Grav.},\ }\textbf {\bibinfo {volume} {26}},\ \bibinfo {pages}
  {155005} (\bibinfo {year} {2009})},\ \Eprint {http://arxiv.org/abs/0810.4193}
  {arXiv:0810.4193 [hep-th]} \BibitemShut {NoStop}%
\bibitem [{\citenamefont {Chongchitnan}\ and\ \citenamefont
  {Efstathiou}(2007)}]{Chongchitnan:2007eb}%
  \BibitemOpen
  \bibfield  {author} {\bibinfo {author} {\bibfnamefont {S.}~\bibnamefont
  {Chongchitnan}}\ and\ \bibinfo {author} {\bibfnamefont {G.}~\bibnamefont
  {Efstathiou}},\ }\Doi {10.1103/PhysRevD.76.043508} {\bibfield  {journal}
  {\bibinfo  {journal} {Phys.Rev.},\ }\textbf {\bibinfo {volume} {D76}},\
  \bibinfo {pages} {043508} (\bibinfo {year} {2007})},\ \Eprint
  {http://arxiv.org/abs/0705.1955} {arXiv:0705.1955 [astro-ph]} \BibitemShut
  {NoStop}%
\bibitem [{\citenamefont {Wiggins}(2010)}]{wiggins2010}%
  \BibitemOpen
  \bibfield  {author} {\bibinfo {author} {\bibfnamefont {S.}~\bibnamefont
  {Wiggins}},\ }\href@noop {} {\emph {\bibinfo {title} {Introduction to Applied
  Nonlinear Dynamical Systems and Chaos}}}\ (\bibinfo  {publisher} {Springer},\
  \bibinfo {year} {2010})\BibitemShut {NoStop}%
\bibitem [{\citenamefont {Matos}\ \emph {et~al.}(2009)\citenamefont {Matos},
  \citenamefont {Luevano}, \citenamefont {Quiros}, \citenamefont
  {Urena-Lopez},\ and\ \citenamefont {Vazquez}}]{Matos:2009hf}%
  \BibitemOpen
  \bibfield  {author} {\bibinfo {author} {\bibfnamefont {T.}~\bibnamefont
  {Matos}}, \bibinfo {author} {\bibfnamefont {J.-R.}\ \bibnamefont {Luevano}},
  \bibinfo {author} {\bibfnamefont {I.}~\bibnamefont {Quiros}}, \bibinfo
  {author} {\bibfnamefont {L.}~\bibnamefont {Urena-Lopez}}, \ and\ \bibinfo
  {author} {\bibfnamefont {J.~A.}\ \bibnamefont {Vazquez}},\ }\Doi
  {10.1103/PhysRevD.80.123521} {\bibfield  {journal} {\bibinfo  {journal}
  {Phys.Rev.},\ }\textbf {\bibinfo {volume} {D80}},\ \bibinfo {pages} {123521}
  (\bibinfo {year} {2009})},\ \Eprint {http://arxiv.org/abs/0906.0396}
  {arXiv:0906.0396 [astro-ph.CO]} \BibitemShut {NoStop}%
\bibitem [{\citenamefont {Wang}\ \emph {et~al.}(2011)\citenamefont {Wang},
  \citenamefont {Chen},\ and\ \citenamefont {Chen}}]{Wang:2011bi}%
  \BibitemOpen
  \bibfield  {author} {\bibinfo {author} {\bibfnamefont {P.-Y.}\ \bibnamefont
  {Wang}}, \bibinfo {author} {\bibfnamefont {C.-W.}\ \bibnamefont {Chen}}, \
  and\ \bibinfo {author} {\bibfnamefont {P.}~\bibnamefont {Chen}},\ }\href@noop
  {} { (\bibinfo {year} {2011})},\ \Eprint {http://arxiv.org/abs/1108.1424}
  {arXiv:1108.1424 [astro-ph.CO]} \BibitemShut {NoStop}%
\end{thebibliography}%

\end{document}